%% file: ms.tex
\documentclass{ws-ijmpd}
\usepackage[super,compress]{cite}

\include{own_latex_commands_i}

\begin{document}

\markboth{S. Rosswog}
{Multi-messenger picture of compact binary mergers}

%
\catchline{}{}{}{}{}
%

\title{The multi-messenger picture of compact binary mergers}

\author{Stephan Rosswog}

\address{Astronomy and Oskar Klein Centre, Stockholm University, AlbaNova,\\
SE-106 91 Stockholm, Sweden\\
stephan.rosswog@astro.su.se}

\maketitle

\begin{history}
\received{Day Month Year}
\revised{Day Month Year}
\end{history}

\begin{abstract}
In the last decade, enormous progress has been achieved in the understanding of
the various facets of coalescing double neutron star and neutron black hole 
binary systems. One hopes that the mergers of such compact binaries can be routinely
detected  with the advanced versions of the ground-based gravitational wave
detector facilities, maybe as early as in 2016.  From the theoretical side, there
has also been mounting evidence that compact binary mergers could be major
sources of heavy elements and these ideas have gained recent observational
support from the detection of an event that has been interpreted as a
``macronova'', an electromagnetic transient powered by freshly produced, radioactively
decaying heavy elements.  In addition, compact binaries are the most plausible
triggers of short gamma-ray bursts (sGRBs) and the last decade has witnessed
the first detection of a sGRB afterglow and subsequent observations have
delivered a wealth of information on the environments in which such bursts occur.
To date, compact binary mergers can naturally explain most --though not all--
of the observed sGRB properties. This article reviews major recent developments
in various areas related to compact binary mergers.
\end{abstract}


\ccode{PACS numbers:}

\section{Introduction}
In July 1974 Russel Hulse and Joseph Taylor discovered the pulsar PSR 1913+16
during a systematic search for new pulsars at the Arecibo Observatory in Puerto
Rico\cite{hulse75}. The pulsar was detected with a period of 59 ms, but with 
apparent changes of 80 $\mu s$ from day to day. Soon it became clear that these
changes are due to the Doppler shift caused by the orbital motion around an
unseen companion star. The orbital period is only 7.75 hours and with a
semi-major axes of about 3 $R_\odot$, but no observed eclipses, the companion 
star must also be a compact object, either a neutron star (ns) of a black hole (bh).
The binary system is so tight that the orbital velocity is of order $10^{-3} c$ and 
general-relativistic effects build up qickly. To quantify the deviation of the dynamics
from a purely Newtonian binary, one introduces  so-called ``Post-Keplerian'' (PK) parameters.
Since these PK parameters each depend on both component masses, one can determine 
{\em individual} stellar masses
through the measurements of at least two PK parameters and not just  --as in Newtonian
theory--  the total binary mass via of Kepler's third law. PSR 1913+16 has 
by now been observed for more than four decades and its parameters have been determined
to an astonishing precision. For example, 
the current values\cite{weisberg10} for the neutron star masses are $m_{\rm p}= 1.4398 \pm 0.0002$ 
\Msun for the pulsar and $m_{\rm c}= 1.3886 \pm 0.0002$ \Msun for the companion star. 
The binary is  a ``clean'' system in the sense that the components can be very accurately
described as point masses and in particular tidal effects can be safely neglected. The periastron 
advances at a rate\cite{weisberg10} of $\dot{\omega}= 4.226598 \pm 0.000005$ deg/yr 
(as compared to 43 arcsec per century for Mercury) and the orbit decays at a rate that agrees
with General Relativity's prediction for  the emission of gravitational waves\cite{maggiore08}
\be
\dot{P}_b= - \frac{192 \pi G^{5/3}}{5 c^5} \frac{m_p m_c}{(m_p + m_c)^{1/3}}
\left( \frac{2 \pi}{P_b}\right)^{5/3} \times \frac{1 + (73/24) e^2 + (37/96) e^4}{(1-e^2)^{7/2}}
\ee
to about 0.3\% \cite{weisberg10}. The latter, however, is only true if the relative Galactic acceleration 
difference between the pulsar and the solar system --they have a separation\cite{weisberg10} 
of about $(9.9 \pm 3.3)$ kpc-- is taken into account. General Relativity had been probed in a number of solar system tests
such as the perihelion shift of mercury\cite{einstein15}, gravitational light deflection\cite{dyson20,froeschle97}
or the Shapiro delay\cite{shapiro64}, or, more recently the Lense-Thirring effect in satellite orbits\cite{ciufolini04,everitt11},
but neither had the predicted radiative properties of gravity been probed nor regimes where the deviation
from a flat space is substantially larger than in the solar system  (gravitational potential $\Phi/c^2\sim 10^{-6}$).
The orbital decay of PSR 1913+16 was the first --though still indirect-- confirmation of the existence
of gravitational waves and it laid to rest a longstanding debate\cite{einstein37,bondi59} about the existence 
of gravitational waves and its quadrupolar nature. In 1993 Hulse and Taylor were awarded the Nobel Prize in 
Physics for their discovery of  PSR 1913+16.\\
In 2003 an even better suited laboratory for strong field gravity\cite{kramer08,kramer09b} was 
discovered\cite{burgay03,lyne04}:
the ``double pulsar'' PSR  J0737-3039 where both components are active pulsars, usually referred to
as pulsar A and B. With an orbital period of only 147 minutes, relativistic effects are even larger and 
with a distance of only about 1 kpc the systematic errors due to the relative Galactic acceleration are 
much smaller. The double pulsar has an eccentricity of $e=0.088$ and shows with
$\dot{\omega}= 16.8991 \pm 0.0001$ deg/yr an even larger periastron advance than PSR 1913+16.
Its masses have again been very accurately measured as $m_A= (1.3381 \pm 0.0007)$ \Msun
and $m_B= (1.2489 \pm 0.0007)$ \Msun\cite{kramer13} and within less than 10 years this binary has become
the best testbed to probe the accuracy of gravitational quadrupole emission with a current accuracy
well below 0.1\%\cite{kramer13}.\\
To date over 2000 pulsars are known, about 10\% of them possess a binary companion\cite{manchester05}
and 10 systems consist of two neutron stars\cite{lorimer08}. Such systems have turned out to be precious
laboratories to probe gravitational theories\cite{damour09,wex14,kramer14}. Although binary evolution suggests that
it would be natural to form also a reasonable fraction of neutron star black hole 
binaries\cite{fryer99a,dominik12,dominik13,dominik14}, to date no such system has been found. They 
should, in a number of respects, have properties similar to double neutron star binaries and we will 
therefore follow the common practice to collectively refer to both types as ``compact binary systems''. \\
Typically,  one distinguishes gravity tests in different regimes: {\em 1) quasi-stationary weak field}, 
i.e. velocities $v \ll c$ and the spacetime is close to Minkowskian everywhere, {\em 2) quasi-stationary strong-field}, 
i.e. velocities $v \ll c$, but the spacetime can show substantial deviations from Minkowskian, {\em 3) highly 
dynamical strong-field}. i.e. $v$ becoming comparable to $c$ and involving a strongly curved spacetime 
and finally the {\em 4) radiation regime} where the radiative properties of gravity are tested. The first regime 
includes solar system tests, the second and fourth regime can be probed by well-separated neutron star binaries 
while regime 3) will be probed by the ground-based gravitational wave detectors such as LIGO, 
VIRGO and KAGRA\cite{abbott09a,harry10,ligo,sengupta10,virgo,somiya12,aVIRGO15}. 
They are expected to see the last inspiral stages ($\sim$ minutes) that are initially well described by 
post-Newtonian methods\cite{blanchet06,futamase07} and the subsequent merger and ringdown phase 
where strong-field gravity, hydrodynamic and nuclear equation of state (EOS) effects from the neutron 
star matter determine the dynamics. For an overview over the numerical modeling of these phases we 
refer to recent reviews\cite{shibata11,faber12,rosswog14c}.\\
Compact binary mergers had also been suggested very early on as the production site for ``rapid-neutron capture''
or ``r-process'' elements\cite{lattimer74,lattimer76,lattimer77}, actually {\em before} the discovery of the Hulse-Taylor
pulsar. Although it may seem a natural idea to form r-process elements in the decompression of the extremely
neutron-rich neutron star material, this idea has for a long time been considered as somewhat exotic. This
perception, however, has changed during the last one and a half decades, mainly because compact binary 
mergers (CBM) were found to robustly produce r-process elements without fine-tuning while core-collapse supernovae 
seem seriously challenged to providing the conditions for at least the heaviest r-process elements. This issue 
will be discussed in more detail in Sec.~\ref{sec:heavy_elements}. The idea has been further boosted by the
recent discovery of an nIR transient in the aftermath of the short GRB 130603B. Such transients from radioactive decays
of freshly produced r-process elements had been a prediction of the compact binary merger model. In particular
the characteristic time scale of $\sim$ one week and a peak in the nIR are consistent with extremely heavy 
elements with large opacities having been produced. This topic will be discussed in more detail in Sec.~\ref{sec:transients}.\\
The merger of a compact binary system is also the most likely trigger of short gamma-ray burst (sGRBs).
The idea that the sGRB phenomenon is caused by compact binary mergers has now stood the
test of three decades of observation and it is considered the most plausible model for the ``engine''
behind sGRBs, although it is not completely free of tension with some observations.
We will review in Sec.~\ref{sec:GRB} the major arguments that link sGRBs with the idea
of an compact binary merger origin, we will discuss how the predictions from the compact binary 
merger model compare to the observed GRB properties and where open questions remain.

In the following, various facets of compact binary mergers are discussed in more detail. We begin by discussing
relevant time scales in Sec.~\ref{sec:time_scales}, followed by the expected neutrino emission in 
Sec.~\ref{sec:neutrino}, nucleosynthesis and closely related issues are treated in Sec.~\ref{sec:heavy_elements} 
and we discuss short GRBs in Sec.~\ref{sec:GRB}. We will conclude with a summary in Sec.~\ref{sec:summary}.

\section{Time scales}
\label{sec:time_scales}
As a stellar binary system revolves around its centre of mass it emits energy and angular 
momentum via gravitational waves, which are both extracted from the orbital motion\footnote{Here, we
will assume that we are dealing with a binary system of two non-spinning point masses
in quadrupole approximation as first worked out by Peters and Mathews in 1963 \cite{peters63}.}
with a power $P \propto \omega_{\rm orb}^{10/3}$. Since the total, kinetic plus potential, orbital
energy is $E_{\rm orb}= -G m_1 m_2/2 a$ a loss of energy means a reduction of the mutual 
separation, which, due to $\omega_{\rm orb}= \sqrt{G M/a^3}$, corresponds to an increase of 
the orbital frequency and therefore to a further enhanced gravitational wave emission. This
runaway process finally leads to a merger of the two stars.
For elliptical orbits, not only the semi-major axis $a$, but also the eccentricity $e$
evolves in time. For a non-zero eccentricity, its the temporal change is negative,
$de/dt< 0$, and the eccentricity is ``radiated away'' very efficiently. The Hulse-Taylor pulsar, 
for example, has currently an eccentricity of $e=0.617$ and a semi-major axis $a=2.2 \times 10^9$ cm. 
By the time the orbit has shrunk to 10 $R_{\rm ns}$, however, it will have\cite{maggiore08} an 
eccentricity of $e<10^{-5}$, i.e. the orbit will be {\em very} close circular.\\
\begin{figure}[]
\centerline{
\psfig{file=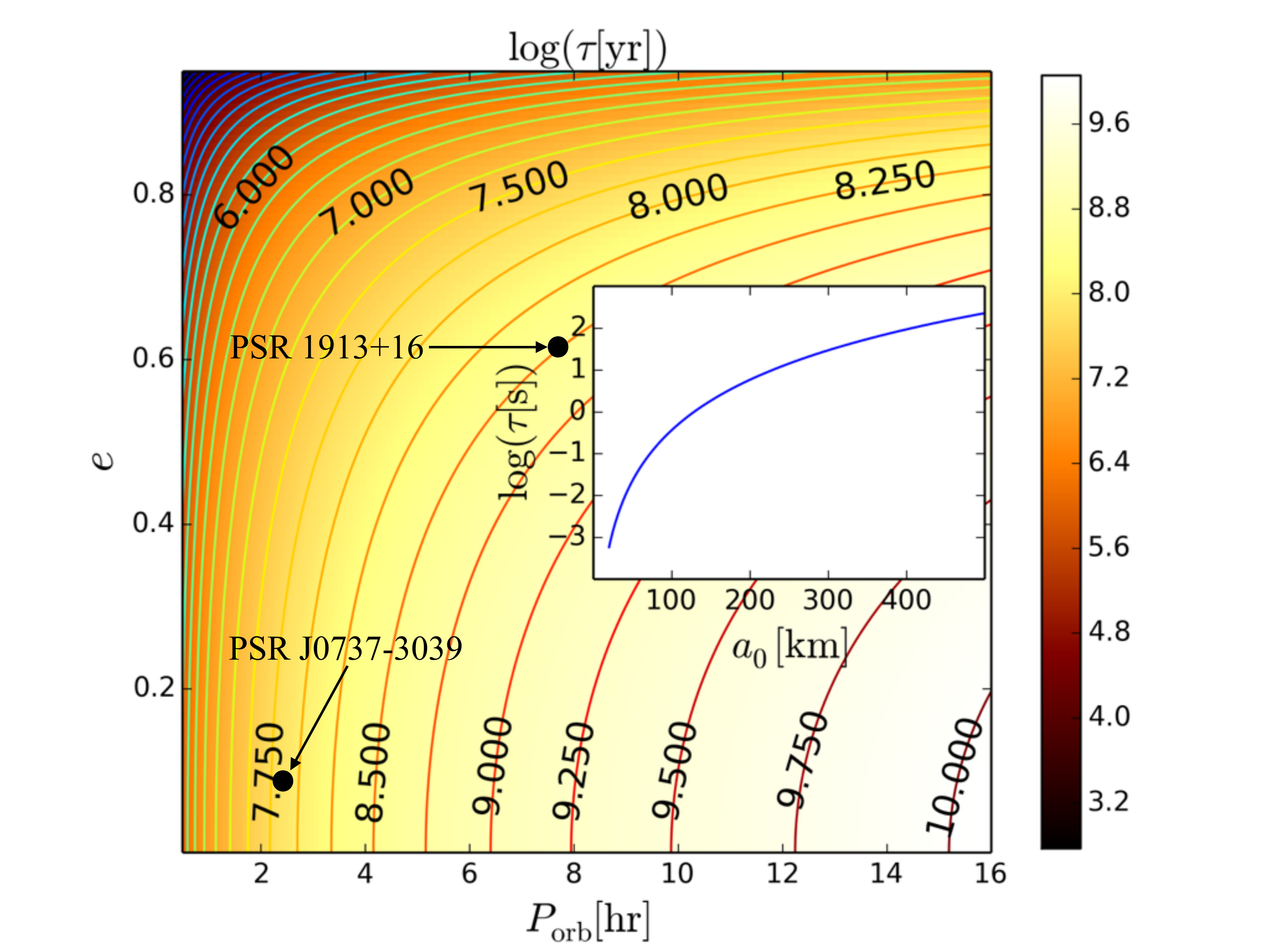,width=13cm,angle=0}
}
\vspace*{0cm}
\caption{Inspiral time for a binary with $2 \times 1.4$ \msun. Shown are contours of $\log(\tau_{\rm GW})$ (in years)
               as function of the current orbital period (in hours) and eccentricity. The filled circles indicate period and 
               eccentricity of the Hulse-Taylor Pulsar PSR 1913+16 and the Double Pulsar J0737-039 (although their
               masses are not exactly 1.4 \msun). The inset focuses on the last inspiral stages where the orbits are
               to very high accuracy circular (note that the axes are different from the main plot). The last stages 
               of the inspiral ($a_0 < 100$ km) only take fractions of a second.}
\label{fig:inspiral}
\end{figure}
A non-zero orbital eccentricity can substantially shorten the inspiral time until coalescence. In
general, the evolution equations for $a$ and $e$ need to be integrated numerically, but a good
approximation for not too large eccentricities\footnote{This estimate ignores an additional eccentricity-dependent
factor which is to within a few percent equal to unity for $e<0.6$, and approaches monotonically $\approx 1.8$
in the limit of $e \rightarrow 1$, so that the estimates are always substantially better than a factor of 2.}
 is given by\cite{maggiore08}
\be
 \tau_{\rm GW}\simeq 9.83 \times 10^6 {\rm years} \left(\frac{P}{\rm hr}\right)^{8/3} \left(\frac{M}{M_\odot} \right)^{-2/3} 
  \left(\frac{\mu}{M_\odot} \right)^{-1} (1-e^2)^{7/2}.
\label{eq:tau_insp}
\ee
Contours of the inspiral time $\log (\tau_{\rm GW})$ for an equal mass binary system with 1.4 \Msun per star
are shown in Fig.~\ref{fig:inspiral}. The inset shows the inspiral time (in seconds) for the last stages of a 
circular binary system as a function of the separation $a_0$.
Also marked are the locations of PSR 1913+16 and of the double pulsar PSR  J0737-3039 although their masses are 
slightly different from 1.4 \msun.
Circular binaries can only merge within the lifetime of  galaxy ($10^{10}$ yrs) if they possess an
initial orbital period below 16 hrs. The Hulse-Taylor pulsar has a life time of $3 \times 10^8$ years until
coalescence. A binary system with the same properties, but an initial eccentricity of $e=0.99$, in contrast, would merge
within only $10^4$ years. This dependence of the inspiral duration on eccentricity also has important implications 
for compact binary mergers as possible sources of r-process elements, see Sec.~\ref{sec:chem_evol}, and for 
their role as central engines of gamma-ray bursts, see Sec.~\ref{sec:GRB}. \\
The dynamics of the inspiral is also responsible for compact binary systems not being spun up to corotation
when the coalescence occurs. For example, it takes only 0.76 s for a $2 \times 1.4$ \Msun binary system 
from the last 120 km of separation ($\approx 10 R_{\rm ns}$) until coalescence. Therefore, the stage where tides 
can be excited is simply too short to enforce corotation, even for an implausibly large internal 
viscosity\cite{bildsten92,kochanek92}. Therefore, the spin period at merger is much smaller than the orbital frequency
and an irrotational binary configuration is a sensible initial condition for modeling a binary system
prior to merger.\\
The frequency of the emitted gravitational waves $f_{\rm GW}$ as a function of the time until coalescence $\tau_{\rm GW}$
is\cite{maggiore08}
\be
f_{\rm GW}= 134 {\rm Hz} \left( \frac{1.21 M_\odot}{M_{\rm chirp}}\right)^{5/8} \left( \frac{1 s}{\tau_{\rm GW}}\right)^{3/8},
\ee
where the ``chirp mass'' is $M_{\rm chirp}= \mu^{5/3} M^{2/5}$. Therefore, a $M_{\rm chirp}= 1.21$ \Msun binary 
would emit gravitational waves of 10 Hz or more (roughly the lowest frequencies accessible to ground-based 
detectors) during the last 17 minutes and 100 Hz or more during the last two seconds of the inspiral. For a 
double neutron star system, one expects the final merger to occur around gravitational wave frequencies of 1 kHz.\\
The inspiral can be accurately described by means of Post-Newtonian methods\cite{blanchet06,futamase07}, 
but the final merger and ``ringdown'' require numerical simulations\cite{shibata11,faber12,rosswog14c}. 
Simulations indicate that the central object that forms in a merger does in many cases not collapse directly 
to a black hole, but instead survives as a ``hypermassive neutron  star'' (HMNS)\cite{baumgarte00,kaplan14,kastaun14,takami14}, despite having 
a mass in excess of  the  Tolman-Oppenheimer-Volkoff maximum mass.
\begin{figure}[pb]
\centerline{\psfig{file=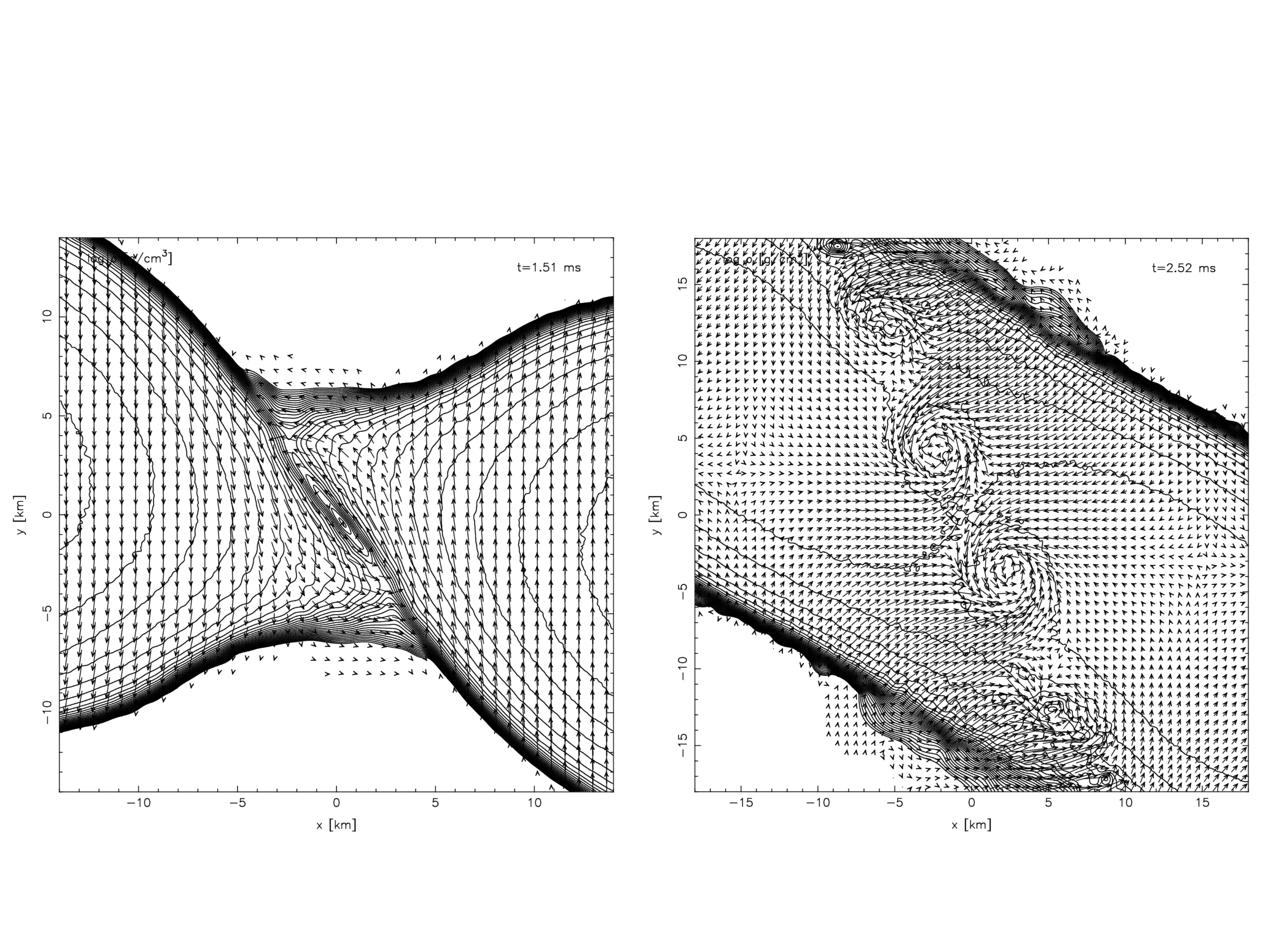,width=12.5cm,angle=0}}
\vspace*{8pt}
\caption{Velocity fields during the merger of two initially non-spinning 1.4 \Msun neutron stars\cite{price06}. 
The shear interface, see panel one, becomes Kelvin-Helmholtz unstable and forms a string of 
vortex rolls. Subsequently the vortex rolls merge and the remnant finally forms a radidly 
differentially rotating, so-called ``hypermassive neutron star'' that is temporarily stabilized 
against the gravitational collapse to a black hole.}
\label{fig:diff_rot}
\end{figure}
This is due to effects such as thermal pressure or --more importantly-- differential rotation\footnote{{\em Differential}
rotation is much more efficient in stabilizing stars than uniform rotation since --in the center-- the rotation can be very rapid without shedding
mass at the surface.}.
A typical velocity field inside the merging binary system is shown in Fig.~\ref{fig:diff_rot}:
at contact a shear interface forms for irrotational binaries (left panel) that becomes Kelvin-Helmholtz unstable
(right panel) and evolves into a differentially rotating remnant. 
Differential rotation is very efficient in stabilizing stars\cite{ostriker68,baumgarte00} and the time scale 
to collapse will be set by angular momentum transport processes that occur on a time scale that is 
substantially longer than the dynamical time scale 
\be
\tau_{\rm ns}= 0.2 \; {\rm ms} \left( \frac{\bar{\rho}}{\rho_{\rm nuc}}\right)^{-1/2},
\label{eq:tau_dyn_ns}
\ee
where $\rho_{\rm nuc}= 2.65 \times 10^{14}$ \Gcc is the nuclear saturation density.
Relativistic simulations of double neutron star mergers \cite{shibata06c,hotokezaka11,hotokezaka13b} indicate 
the central remnant is a HMNS\footnote{In the following, we will use the abbreviation HMNS collectively for both
``hypermassive neutron stars'' that will finally collapse into a black hole and also for very massive, but stable, 
neutron stars.}, unless the initial ADM mass of the binary exceeds a threshold mass of 1.35 times 
the maximum mass of a cold, non-rotating neutron star. The discovery of two massive neutron stars, J1614-2230 with 
$M_{\rm ns}= 1.97 \pm 0.04$ \Msun \cite{demorest10} and PSR J0348+0342 with $M_{\rm ns}= 2.01 \pm 0.04$ \Msun 
\cite{antoniadis13}, now place this limit to $M_{\rm thresh} > 2.7$ \Msun so that --depending on the distribution of 
neutron star masses that is realized in binary systems-- a large fraction of the mergers may go through such 
a metastable phase. Taking the current mass estimates for binary neutron stars\cite{lattimer12a} at face value 
(but keep in mind that some estimates have substantial errors) only three out of the 
10 DNS systems could avoid such a metastable phase and would instead directly collapse to a black hole. 
The delay time between merger and bh formation
also has a crucial impact on both the triggering of short GRBs and --since it substantially impacts 
on neutrino-driven winds-- on the merger nucleosynthesis. These issues
are further discussed in Sec.~\ref{sec:heavy_elements} and \ref{sec:GRB}. A long-lived HMNS could
also produce additional radiative sigatures and contributions to high energy cosmic rays\cite{yu13,takami14c,metzger14b}.\\
In most cases a CBM leads to the formation of a ``thick'' accretion disk where the scale height $H$ is comparable
to the radius $R_{\rm disk}$\footnote{We do not distinguish here between ``disk'' and ``torus'' and use both 
words synonymously.}. The dynamical time scale of such a disk is roughly
\be
\tau_{\rm dyn, disk}\sim \frac{2 \pi}{\omega_{\rm K}} \approx 0.01 s\left(\frac{M}{2.5 M_\odot} \right)^{-1/2} 
\left( \frac{R_{\rm disk}}{100 \; \rm km}\right)^{3/2},
\label{eq:tau_dyn_disk}
\ee
while its viscous time scale is
\be
\tau_{\rm visc} \sim 0.3 \; {\rm s} \;  \left(\frac{0.05}{\alpha}\right)  \left(\
 \frac{R_{\rm disk}}{100 {\rm \; km}} \right)^{3/2}
\left(\frac{2.5 M_\odot}{M}\right)^{1/2} \left( \frac{R_{\rm disk}/H}{3} \right)^2,
\label{eq:tau_visc}
\ee
where we have assumed that the viscosity can be parametrized as a Shakura-Sunyaev-type dissipation\cite{shakura73}.


\section{Neutrino emission}
\label{sec:neutrino}
Like supernovae, compact binary mergers emit a large fraction of the released gravitational binding energy
($\sim 10^{53}$ erg) in the form of neutrinos. This results predominantly in neutrinos in the energy range of 
$\sim 20$ MeV. Their moderate energies together with the steep energy dependence of the interaction cross 
sections make them hard to detect\footnote{Keep in mind that so far neutrinos have been detected only once
  for a supernova\cite{arnett89}, SN1987A, and compact binary mergers occur $\sim 10^3$ times less frequently.}.
But if the merger also accelerates material to relativistic speeds, as expected if they are the 
sources of short GRBs (Sec.~\ref{sec:GRB}), one may also expect neutrinos of substantially larger energy. 
The same shocks that are thought to accelerate the electrons responsible for the prompt gamma-ray 
emission should also produce relativistic protons, which can produce high-energy 
neutrinos\cite{waxman04a,waxman04b,dermer05,waxman06a}. 
Internal shocks that can produce the observed $\sim$ MeV photons, should also be able to produce 
so-called prompt neutrino emission with energies in excess of $10^{14}$ eV \cite{waxman97,rachen98,waxman99,dermer03}. 
High-energy neutrinos may also be produced via neutron-rich outflows \cite{derishev99,bahcall00} and reverse 
shocks\cite{waxman00}.
Such neutrino production mechanisms, however, rely on GRB outflows having a sizeable baryonic component. 
If instead the outflow should be essentially Poynting flux, large neutrino fluxes would be unexpected, unless 
they are produced efficiently  in the forward shock \cite{dermer02,li02}.\\
Our main focus here, however, is the  lower neutrino energy channel coming  directly from the merger remnant.
Until merger, the neutron stars are still in cold $\beta$-equilibrium, since tidal dissipation causes only a raise 
to moderate temperatures\cite{lai94c} ($T\sim 10^8$ K) and the tidal interaction has practically no time to change the 
neutron-to-proton ratio by weak interactions, see the inset in Fig.~\ref{fig:inspiral} for an illustration. The 
$\beta$-equilibrium condition on the chemical potentials is given as\footnote{We discuss here only the simplest 
case where neutrons, protons and electrons are present in the neutron star. When the Fermi energies become large 
enough to allow for new particle species to appear, the condition has to be modified accordingly, see for example 
Glendenning\cite{glendenning00}.} 
$\bar{\mu}_n= \bar{\mu}_p + \bar{\mu}_e + \bar{\mu}_{\bar{\nu}_{e}}$, where the bar  indicates 
that the particle mass contributions are included. Since in old neutron stars no neutrinos are present (their
diffusion time is of order seconds, see below) the neutrino chemical potential vanishes and the difference 
between the nucleon chemical potentials equals the electron chemical potential which, in turn, determines 
the electron fraction $Y_e$ which has typical values below 0.1. After the merger, we can
expect to have a fraction of the virial temperature, $T_{\rm vir} \sim 25$ MeV ($M$/2.5 \msun) (100 km/$R$), i.e. 
temperatures substantially beyond the electron-positron pair production threshold ($m_e c^2= 0.511$ MeV). 
Therefore, positron captures $n + e^+ \rightarrow p + \bar{\nu}_e$ drive the electron fraction to higher values 
and yield the copious emission of electron-type anti-neutrinos which are the dominant neutrino 
species. This is different from the core-collapse SN case that yields neutrino luminosities in  a similar regime, 
but dominated by the $\nu_e$ from the neutronization process $p + e \rightarrow n + \nu_e$. In the hot, 
high-density interior of a HMNS one also expects  heavy lepton neutrinos to be 
produced\cite{ruffert97a,rosswog03a,dessart09,perego14b}. Since they can only interact via neutral 
current reactions, they escape easier from hotter and denser regions and therefore possess higher energies.
Clearly, for the conditions prevailing in the remnant ($\rho > 10^{14}$ gcm$^{-3} $ in the HMNS and $\sim 10^{10}$ 
gcm$^{-3} $ in the disk) photons are completely trapped on the relevant time sacles and neutrinos are the only cooling 
agents.\\
In the following, we will assume that a HMNS with temperature $T_{\rm HMNS}= 20$ MeV, a mass $M= 2.5$ \Msun and with 
a radius $R=20$ km is present, so that its average density is $\bar{\rho}\approx 1.5 \times 10^{14}$ \gcc ($M$/2.5 \msun) 
($R$/20 km)$^{-3}$. We further assume for our estimates the presence of a surrounding accretion disk of  $T_{\rm disk}= 5$ MeV,
$M_{\rm disk}= 0.2$ \msun, a typical radius $R_{\rm disk}=100$ km, a typical density $\rho_{\rm disk}= 10^{11}$ \Gcc 
and an aspect ratio $H/R= 1/3$, where $H$ is the characteristic disk height. 
To obtain an order of magnitude estimate for the neutrino properties, we assume for simplicity that the main 
source of opacity is provided by neutrinos scattering off nucleons\footnote{For $\nu_e$ the opacity related to 
the absorption by neutrons is even larger, but still of the same order.} with a cross-section given by\cite{tubbs75}
$\sigma = (1/4) \sigma_0 (E_\nu/m_e c^2)^2$, where the reference cross-section is $\sigma_0= 1.76 \times 10^{-44}$ cm$^2$. 
Therefore, the mean free path\footnote{The expected neutrino energies are moderate multiples of the matter 
temperature, usually given by the ratio of two Fermi-integrals at the local chemical potential. For our scalings we use  
$E_\nu \approx 3 k_B T$, i.e. 60 MeV for the HMNS and 15 MeV for the disk.} in the HMNS is 
\be
\lambda_{\nu}^{\rm HMNS}= \frac{1}{n \sigma}  \approx 1.8 \; {\rm m} \left( \frac{R}{20 \; \rm km}\right)^{3} \left( \frac{M}{2.5 \; M_\odot} \right)^{-1} \left(\frac{E_{\nu}}{60 \; \rm MeV} \right)^{-2}
\ee
and the corresponding neutrino optical depth is\footnote{We abbreviate both the optical depth and typical 
time scales with the symbol $\tau$. Since each time it has unique subscript this should no lead to confusion.}
\be
\tau_{\nu}^{\rm HMNS} \sim \frac{R}{\lambda_{\nu}^{\rm HMNS}} \sim 10^4 \left( \frac{R}{20 \; \rm km}\right)^{-2}  \left( \frac{M}{2.5 \; M_\odot} \right) \left(\frac{E_{\nu}}{60 \; \rm MeV} \right)^2
\ee
i.e. the HMNS is very opaque to its own neutrinos.
The characteristic neutrino diffusion time scale is then\cite{ruffert97a,rosswog03a}
\be
\tau_{\rm diff}^{\rm HMNS} \sim 3 \tau_{\nu}^{\rm HMNS} \frac{R}{c} \sim 2.4 \; {\rm s} \left( \frac{R}{20 \; \rm km}\right)^{-1} \left( \frac{M}{2.5 \; M_\odot} \right) \left(\frac{E_\nu}{60 \; \rm MeV} \right)^2, 
\label{eq:nu_diff_time_ns}
\ee
i.e. on the HMNS dynamical time scale, see Eq.~(\ref{eq:tau_dyn_ns}), the neutrinos are efficiently
``trapped''. If we assume that a thermal energy of $\Delta E_{\rm th}\sim 0.1 G M^2/R$ can in principle be emitted, one expects a HMNS luminosity of
\be
L_\nu^{\rm HMNS}\sim \frac{\Delta E_{\rm th}}{\tau_{\rm diff}^{\rm HMNS}} \sim 3 \times 10^{52} \; {\rm erg/s} \; \left( \frac{M}{2.5 \; M_\odot} \right) \left(\frac{E_\nu}{60 \; \rm MeV} \right)^{-2},
\ee
and, applying similar estimates as before, one finds the mean free path in the disk
\be
\lambda_{\nu}^{\rm disk} \approx 42 {\rm km} \left( \frac{\rho}{10^{11} \rm g cm^{-3}}\right)^{-1}  \left( \frac{E_\nu}{15 \rm MeV}\right)^{-2}
\ee
and therefore the optical depth is $\tau \sim H/\lambda_{\nu}^{\rm disk} \sim 1$. Thus, the typical escape time for a neutrino
is substantially shorter than  the dynamical and viscous time scales of the disk, see Eqs.~(\ref{eq:tau_dyn_disk}) and (\ref{eq:tau_visc}). 
This short cooling time scale implies that neutrino emission can only be maintained through the constant supply of thermal energy via 
accretion. The gravitational energy gained by the accretion of the disk is then $\Delta E \sim G M M_{\rm disk}/R$ and the expected neutrino
luminosity becomes
\bea
L_{\nu}^{\rm disk} &\sim& \frac{1}{2} \frac{\Delta E}{\tau_{\rm visc}} \nonumber\\
&\sim& 10^{53} \frac{\rm erg} {\rm s} 
\left( \frac{M}{2.5 M_\odot}\right)^{\frac{3}{2}}  
\left( \frac{M_{\rm disk}}{0.2 M_\odot}\right)
\left( \frac{\alpha}{0.05}\right)
\left( \frac{R_{\rm disk}}{3H} \right)^{2}  
\left( \frac{100 \; \rm km}{R_{\rm disk}} \right)^{\frac{3}{2}}  
\left( \frac{20 \; \rm km}{R}\right).
\eea
\\
Clearly, these are only order of magnitude estimates, for more precise results numerical simulations are required. 
To date only a few implementations of neutrino physics in merger simulation codes 
exist\cite{ruffert97a,rosswog03a,sekiguchi10b,deaton13a,neilsen14}. It is encouraging that despite the steep
temperature dependence of the neutrino emission rates (the electron capture energy emission rates, for example
\cite{bruenn85}, are $\propto T^6$), the predicted neutrino luminosities\cite{ruffert01,rosswog03a,sekiguchi11,rosswog13a,neilsen14} 
are rather robust, for a $2 \times 1.4$ \Msun system around $2 \times 10^{53}$ erg/s with average energies
around 10, 15 and 20 MeV for $\nu_e$, $\bar{\nu}_e$ and $\nu_x$, respectively, and, as expected from 
the simple considerations, with the luminosity being dominated by $\bar{\nu}_e$. It seems in particular that 
implementation details play less of a role than the physical ingredients such as the treament of gravity or the 
equation of state. While the quoted results are based on relatively simple leakage schemes, more sophisticated 
neutrino treatments\cite{dessart09,perego14b} yield actually similar results. It had however, been realized by 
Dessart et al.\cite{dessart09} that nucleon-nucleon bremsstrahlung, which had been ignored in early 
implementations, is also an important source of heavy lepton neutrinos.\\
More recently, neutrino emission has also been included in nsbh-merger simulations \cite{deaton13a,foucart14}.
Here the parameter range, especially the expected mass ratio range, is in principle even larger (and to date not yet
well explored), but the first studies\cite{foucart14} find, for $M_{\rm bh}= 7$ \Msun and large bh spin ($a_{\rm bh}>0.7$), 
also luminosities around $\sim 2\times 10^{53}$ erg/s. Lower mass black holes are expected to yield hotter disks
and therefore larger neutrino luminosities. Indeed, a simulation for a 5.6 \Msun bh with $a_{\rm bh}= 0.9$ yield
peak luminosities $> 10^{54}$ erg/s. The bh mass distribution in nsbh binaries, however, is not well known
and a recent study\cite{oezel10}  based on X-ray binaries seems to suggest a peak around $\sim 8$ \msun.
These results are consistent with studies\cite{belczynski08} that find, from a binary evolution point of view,
bh masses near $\sim 10$ \Msun most likely.

\section{Compact binary mergers as cosmic factories of the heaviest elements}
\label{sec:heavy_elements}
The heaviest nuclei in the Universe form via neutron captures. In this process there is a competition
between captures of new neutrons and $\beta-$decays and the ratio of the two time
scales can be used to distinguish between a ``slow'' (s-process) and a ``rapid neutron capture process''
(r-process)\cite{burbidge57,cameron57b,cowan91,woosley92,arnould07}. Each process is responsible for about half 
of the elements heavier than iron. Although the physical mechanisms have been understood decades ago, 
there is still no concensus about the astrophysical production site of the r-process. For many 
years supernovae have been considered as the most promising site, in particular the neutrino-driven 
wind that emerges after the formation of a hot, still deleptonizing proto-neutron 
star\cite{duncan86,qian96b,hoffman97,freiburghaus99a,otsuki00,thompson01,farouqi10}. Over the last decade, 
however, via increasingly sophisticated supernova simulation models\cite{arcones07,fischer10,huedepohl10,roberts10} 
the concensus has grown that the neutrino-driven winds do not provide the right physical conditions 
for the production of the third r-process peak ($A \sim 195$), but contributions to the less heavy region up 
to the second peak near $A \sim 130$ may be plausibly expected, see Arcones and Thielemann\cite{arcones13a} for
an excellent review. A possible 
alternative supernovae site could be magnetohydrodynamically launched jets\cite{winteler12b}, but 
these require a rather exreme combination of spin and magnetic field as initial conditions and at present
it is not clear whether or at which rate such conditions are realized in nature.\\
Neutron-rich matter ejected in a compact binary merger has been considered
as an alternative formation channel of rapid neutron capture ("r-process") 
elements since the 70ies\cite{lattimer74,lattimer76,lattimer77,eichler89}. Despite its physical plausibility,
this channel was considered exotic for a long time and only the second-best model after core-collapse
supernovae. This perception has only changed in the last one and a half decades
through a number of studies\cite{rosswog99,freiburghaus99b,ruffert01,oechslin07a,roberts11,goriely11a,korobkin12a,wanajo14,just14}
that found r-process, up to the third r-process peak ($A \sim 195$), naturally occuring in compact binary
merger ejecta.
The ease with which r-process is produced by compact binary mergers together with the serious difficulties 
to produce heavy r-process in core-collapse supernovae \cite{arcones07,fischer10,huedepohl10,roberts10} has
lead to a shift in the general opinion towards compact binary mergers as the most likely production site of
at least the heaviest, but possibly of even all of the r-process elements. An important issue that needs to be
better understood, however, is whether/how r-process from compact binary mergers is consistent with 
galactochemical evolution.\\

\begin{figure}[pb]
\centerline{\psfig{file=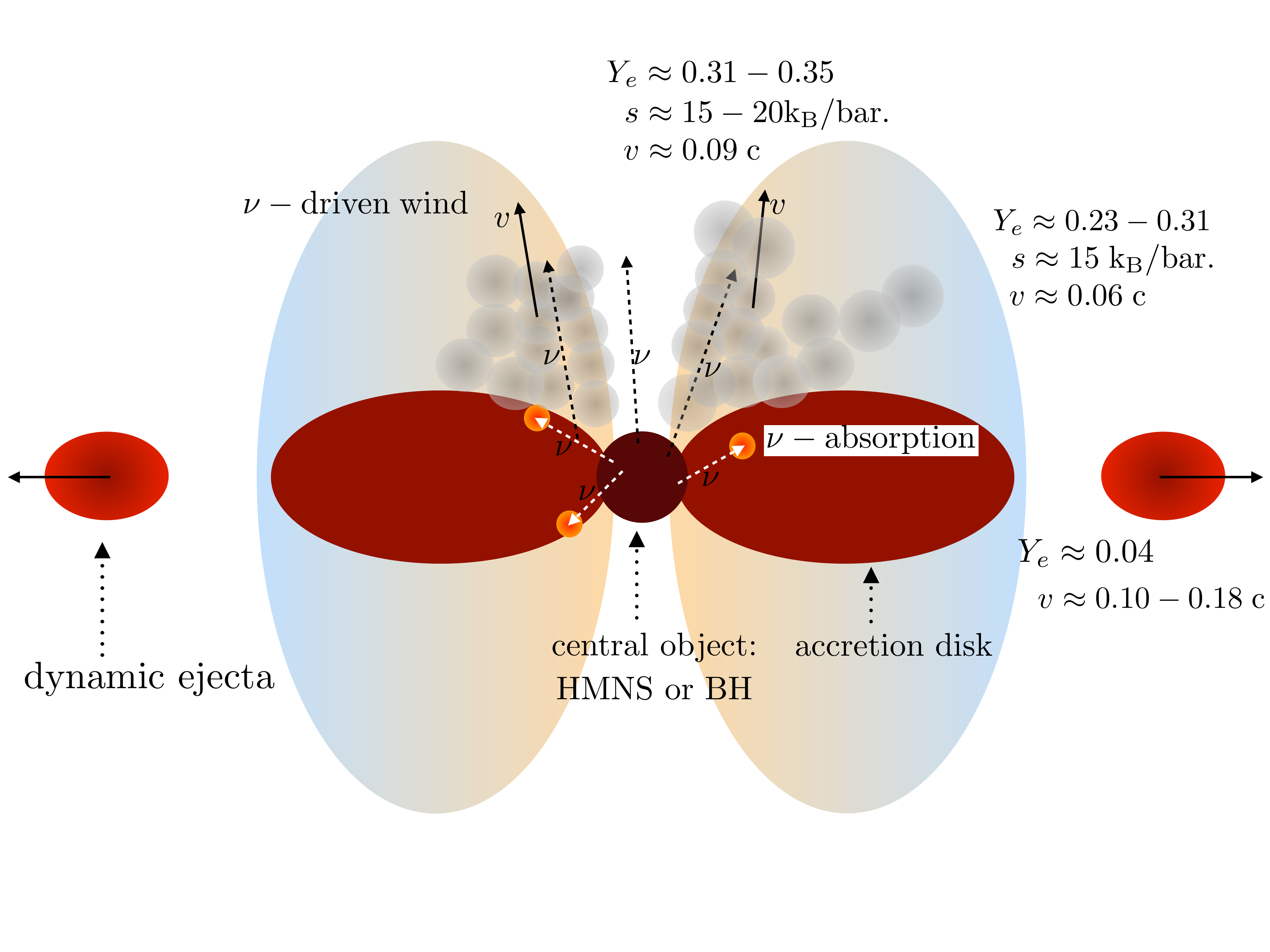,width=10cm,angle=0}}
\vspace*{8pt}
\caption{A schematic illustration of the merger remnant. The numerical values are from recent numerical
               studies\cite{rosswog13a,rosswog13b,perego14b}.}
\label{fig:sketch_remnant}
\end{figure}

\subsection{Mass loss chanels}
\label{sec:mass_loss_channels}
During the merger of a compact binary system mass is ejected into space via a number of channels.
A fraction is released dynamically, i.e. via hydrodynamic interaction and gravitational torques, and will 
subsequently be referred to  as ``dynamic ejecta''. Most CBMs also result in an 
accretion disk, and as such a disk evolves on a viscous time scale, a substantial fraction of its initial 
mass becomes unbound as a result of nuclear and viscous 
action\cite{metzger08a,beloborodov08,lee09,fernandez13a,fernandez13b,just14}. Another channel
that has received a fair amount of attention recently are neutrino-driven winds 
\cite{ruffert97a,rosswog02b,rosswog03b,rosswog03c,dessart09,perego14b,just14}.
Highly magnetized neutron star-like merger remnants can also magnetically drive 
winds\cite{shibata11b,kiuchi12b,siegel14a,kiuchi14}. Also high-velocity ejecta coming from
the interaction region between two neutron stars have been postulated\cite{kyutoku14,metzger15}. 
All these channels differ in the amount of ejected mass, the distribution of electron fractions and entropies.
Therefore, they produce likely different element distributions and --if all produce electromagnetic transients--
these may be different for each channel.
\\
What is an ejecta amount that is interesting from a galactic r-process enrichment perspective?
If $\mathcal{R}_i$ denotes the rate of event type $i$ (averaged over the Galactic age) and
$\bar{m}_{i,c}$ is the average ejected mass per event in a particular channel $c$ of this event type, then the 
average Galactic r-process enrichment rate is
\be
\dot{\mathcal{M}}_{\rm r, gal} = \frac{M_{\rm r, gal}}{\tau_{\rm gal}} = \sum_i  \mathcal{R}_i \sum_c \bar{m}_{i,c} 
\equiv \sum_i  \mathcal{R}_i \bar{m}_{i},
\ee
where $\dot{\mathcal{M}}_{\rm r, gal} $ is the r-process enrichment rate, averaged over 
the age of the Galaxy $\tau_{\rm gal}$. In the following estimates we will conservatively restrict our considerations
to nucleon numbers $A > 140$, since a large number of studies has shown that dynamic ejecta
robustly produce such heavy r-process elements. It is, however, also a possibility that CBM produce
all  {\em all} r-process material\cite{wanajo14,just14} and it is also possible that event types different
from CBMs contribute to the r-process production. This latter possibility will be ignored in our below estimates,
therefore the quoted numbers should be interpreted as upper limits.\\
If we assume that the solar abundance pattern\cite{kaeppeler89} is representative for the Milky Way, 
we can multiply the mass fraction of nuclei\cite{sneden08} with $A>140$, $X_{\rm r}^{>140}= 2.6 \times 10^{-8}$, with the 
baryonic mass of the Milky Way\cite{mcmillan11a}, $6 \times 10^{10}$ \msun, to find a Galactic r-process
mass ($A>140$)  of $M_{\rm r, gal}^{>140} \approx 1560$ \Msun and therefore the average r-process enrichment rate ($A>140$) is
$\dot{\mathcal{M}}_{\rm r, gal}^{>140} \approx 1.1 \times 10^{-7}$ \Msun year$^{-1}$.
Since nsbh mergers are estimated to be at least an order of magnitude rarer than nsns mergers\cite{abadie10},
we only consider the latter for the following order of magnitude estimate\footnote{Using simulation results on 
ejected mass\cite{bauswein14b}, the nsbh rate has recently been constrained to $\mathcal{R}_{\rm nsbh} < 60$ 
MWEG$^{-1}$  Myr$^{-1}$.}. Using "plausible pessimistic" and "plausible optimistic" rates\cite{abadie10} as brackets, 
$\mathcal{R}_{\rm nsns}^{\rm low}\sim 1$ MWEG$^{-1}$  Myr$^{-1}$  and $\mathcal{R}_{\rm nsns}^{\rm high}\sim 1000$ 
MWEG$^{-1}$  Myr$^{-1}$, where MWEG refers to  ``Milky Way-equivalent Galaxy'',  we see that  
\be
\bar{m}_{\rm nsns} \sim \dot{\mathcal{M}}_{\rm r, gal}^{>140} / \mathcal{R}_{\rm nsns} \sim 10^{-4} \; ... \; 0.1 \; M_\odot
\label{eq:relevant_mass}
\ee
is an interesting ejecta amount from a chemical evolution perspective. If instead of only nucleon numbers 
$A>140$ {\em all} r-process elements should be produced in CBM, the ejecta masses would need to be  larger 
by a factor of $X_{\rm r}/X_{\rm r}^{>140} \approx 13$, if there are substantial contributions from other events the
nsns yields need to be correspondingly lower.  These estimates are what hydrodynamic simulation results 
need to be compared with. It further needs to be understood whether such masses/rates are consistent with 
the observed chemogalactic evolution.

\subsubsection{Dynamic ejecta}
In fact, essentially all recent binary merger calculations find dynamic ejecta masses in a range that is consistent with
being a mjor source of cosmic r-process, see Table \ref{tab:ejecta_masses}.
\begin{table}[ph]
\tbl{Comparison of the masses for the dynamic ejecta found in different numerical studies. While the numbers give 
a good impression about the expected mass range, it is worth keeping in mind that they are of limited comparability
since the studies vary different parameters. CF: conformal flatness approximation, GR: General Relativity, PW: 
Paczynski-Wiita potential.}
{\begin{tabular}{@{}lcccc@{}} \toprule
authors & reference & dyn. ejecta [0.01 \msun]  & binary & comment\\
\colrule
Bauswein et al. 2013     & \cite{bauswein13a}    &  $ 1.7 \times 10^{-2}$ ... 1.8 &  nsns  & CF-approximation\\
Bauswein et al. 2014     & \cite{bauswein14b}     & $<2 \times 10^{-4}$ ... 9.6   & nsbh   & CF-approximation\\
Deaton et al. 2013        & \cite{deaton13a}          &  8                                         & nsbh   &  M$_{\rm bh}= 5.6$ M$_\odot$, $a_{\rm bh}= 0.9$  \\
Foucart et al. 2013        & \cite{foucart13}           & 1 ... 5                                    & nsbh  & M$_{\rm bh}= 10$ M$_\odot$, $a_{\rm bh}= 0.9$ \\
Foucart et al. 2014        & \cite{foucart14}           & \hphantom{0}5 ... 20           & nsbh & M$_{\rm bh}= 7 ... 10$ M$_\odot$, $a_{\rm bh}= 0.7 ... 0.9$ \\
Hotokezaka et al. 2013 & \cite{hotokezaka13a}  & 0.01 ... 1.4                            & nsns   &  full GR\\
Kyutoku et al. 2013       &\cite{kyutoku13}           & 1 ... 7                                    & nsbh   &  full GR    \\
Oechslin et al. 2007      & \cite{oechslin07a}       &    0.1 ... 4.5                           & nsns   &   CF-approximation\\
Rosswog et al. 2013     & \cite{rosswog13b}       & \hphantom{0}0.76 ... 3.9       & nsns  &   Newt.\\
Rosswog 2005              & \cite{rosswog05a}       & \hphantom{0}1 ... 20             & nsbh & M$_{\rm bh} \ge 14$ M$_\odot$, PW-potential \\
 \botrule
\end{tabular} \label{tab:ejecta_masses}}
\end{table}
For nsns systems, these dynamic ejecta can be further split into an ``interaction component'' that emerges from the interface between the
two neutron stars and a ``tidal component''\footnote{For an example see Fig. 2 in Korobkin et al. 2012\cite{korobkin12a}.} that is launched by gravitational torques without experiencing noticeable shocks or shear 
flows\cite{oechslin07a,korobkin12a}. The matter from 
the interaction region is considerably hotter than the tidal component. Therefore, it may, via positron captures, change its electron fraction
$Y_e$, while the tidal component is ejected at its low, initial $\beta$-equilibrium value. The ratio of the mass in these components also
depends on the dynamic evolution at merger and therefore on the treatment of gravity and the used equation of state. In Newtonian calculations
\cite{korobkin12a,rosswog13a,rosswog13b} with the stiff Shen et al. EOS\cite{shen98a,shen98b} the tidal component dominates while 
simulations using the Conformal Flatness approximation to GR \cite{bauswein13a} and simulations using dynamical, general-relativistic 
space-times \cite{hotokezaka13a} are dominated by the interaction component. Due to their possibly different electron fractions this
may have implications for nucleosynthesis, see below.

\subsubsection{Disk dissolution}
When an accretion disk forms in the aftermath of a compact binary merger, it is characterized by 
$\rho\simeq10^{10}$ \Gcc, $T \simeq 10^{10}$ K and initial electron fractions close to the 
$\beta$-equilibrium value, $Y_e < 0.1$, of the disrupted neutron star, but increasing as the disk 
evolves viscously. Under such conditions the bulk of the disk consists of free 
neutrons and protons and $e^+ e^-$-pairs. As the disk evolves viscously on a secular time scale, see 
Eq.~(\ref{eq:tau_visc}), it expands and cools and at some point  the free 
nucleons combine into light elements, predominantly alpha particles\footnote{Keep in mind that 
on the viscous time scale weak interactions may have substantially shifted the electron fraction 
$Y_e$ to larger values.}, which releases $\approx 7$ MeV/nucleon. Together with vsicous dissipation,
turbulent energy transport and energy deposition by neutrinos this can unbind a substantial fraction (up to
25 \%) of the original disk mass\cite{metzger08a,beloborodov08,lee09,fernandez13a,metzger14a,just14}. 
How much matter becomes unbound by neutrino absorption is crucially dependent on the the presence
of a central HMNS vs a black hole, in the latter case substantially less material is blown off from the disk
\cite{dessart09,fernandez13a,fernandez13b,perego14b,just14}.
 
\subsubsection{Neutrino-driven winds}
\label{sec:nu_winds}
When two neutron stars merge, they release a substantial fraction of the gravitational binding energy in the 
form of neutrinos. As shown in Sec.~\ref{sec:neutrino}, the remnant neutrino luminosities are in the range of a few times $10^{53}$ erg/s 
with typical neutrino energies of $\sim 15$ MeV\cite{ruffert97a,rosswog03a,sekiguchi11,kiuchi12,neilsen14}. 
The gravitational binding energies per nucleon in the merged remnant, in comparison, are of order 
$E_{\rm grav} \approx - 35 \; \rm{MeV} \; (M_{\rm co}/2.5$ \msun) $(100$ km/$r$), where  $M_{\rm co}$ is the 
mass of the central object. That means capturing a few neutrinos 
provides enough energy to potentially escape the gravitational attraction of the remnant. It had been 
realised early on \cite{ruffert97a,rosswog02b,rosswog03b} that the absorption of neutrino energy could 
drive strong baryonic winds, similar to the case of a new-born proto-neutron star\cite{duncan86,qian96b}, 
see Fig.~\ref{fig:sketch_remnant} for a simple sketch, but the wind properties (mass, electron fraction, entropy 
or geometry) are difficult to estimate without numerical simulations.\\
Early investigations used either order-of-magnitude approaches or  parametrized models to explore the neutrino-winds
from merger remnants
\cite{ruffert97a,rosswog02b,rosswog03c,mclaughlin05,surman06,surman08,metzger08a,wanajo12,caballero12}.
Due to the involved technical challenges, full-fledged numerical simulations of neutrino-driven winds have only been
performed recently. The first neutrino-hydrodynamic study was performed in 2D by Dessart et al. \cite{dessart09}, 
followed by recent studies by Fernandez and Metzger\cite{fernandez13b,metzger14a} and by Just et al.\cite{just14}. 
Just et al. explored the different ejecta channels using viscous hydrodynamics with Newtonian and pseudo-Newtonian
gravity and they applied an energy-dependent two-moment closure scheme for the transport of electron neutrinos and
anti-neutrinos. For black hole disk systems, they found neutrino interactions to help unbind disk material ($\sim 1\%$
of the initial disk mass), however, to a much smaller  extent than viscous action which is able to unbind up to 25\% 
of the initial disk mass. This result is consistent with earlier studies in 2D \cite{fernandez13b} that also found that 
neutrino heating in bh-disk systems unbinds only a small fraction of the disk material.\\
To date, we are aware of only a single study in 3D, performed by Perego et al. \cite{perego14a}.  They 
explored the case where a HMNS is present and used the end points of 3D SPH simulations\cite{price06} as initial 
condition. The further evolution was followed by means of an Eulerian hydrodynamics code\cite{kaeppeli11} augmented 
by a detailed, spectral leakage scheme that has been gauged at Boltzmann transport calculations of supernovae.
They found a strong baryonic wind being blown out along the original binary rotation axis within $\sim 100$ ms,
in qualitative agreement with the earlier 2D work by Dessart et al.\cite{dessart09}. 
This wind unbinds at least $3.5 \times 10^{-3}$ \Msun and is therefore of relevance for the galactic chemical evolution,
see Eq.~(\ref{eq:relevant_mass}). The weak interactions in the wind yield a broad distribution of electron fractions between
$0.2$ and $0.4$, with tendentially more proton-rich outflow ($Y_e >0.3$) along the rotation axis while the equatorial
outflow remained more neutron-rich ($0.2 < Y_e < 0.3$).

\subsection{Nucleosynthesis in different channels}
\label{sec:nucleo_channels}
\begin{figure}[htb]
    \vspace*{-3cm}
         \centerline{\includegraphics[width=9.5cm,angle=-90]{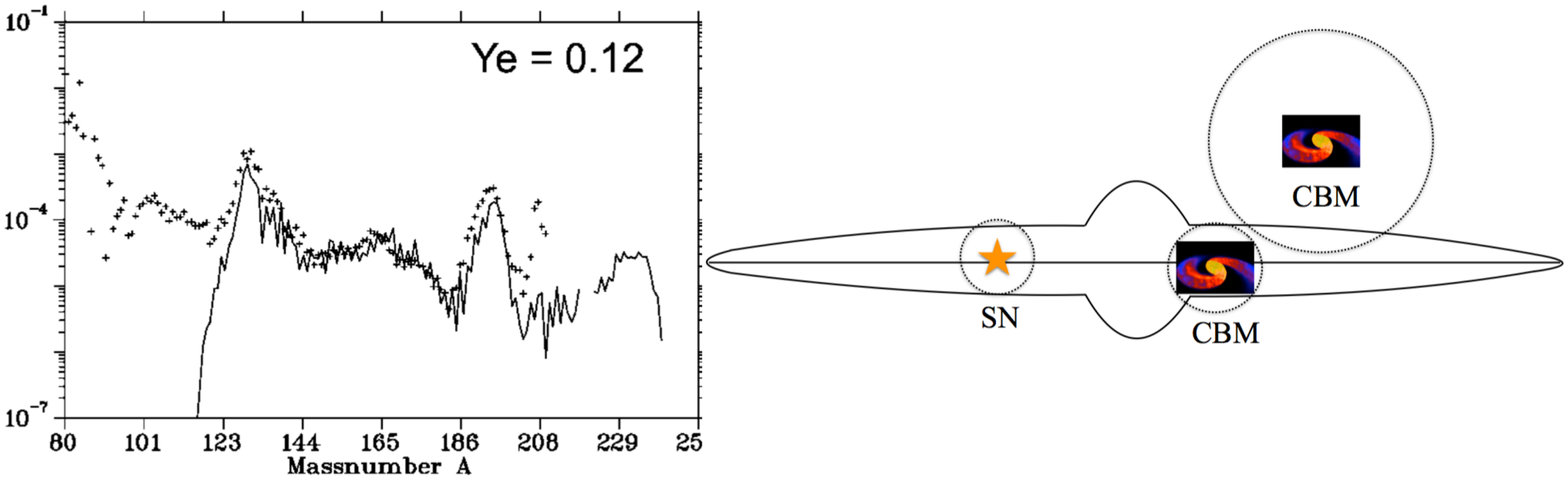}}
    \vspace*{-2cm}
    \caption{Left panel: r-process abundance pattern found the dynamic ejecta from neutron star mergers\cite{freiburghaus99b}
                  are shown as line, the solar system abundance pattern as crosses. Right panel: sketch of different event locations
                  with respect to their host galaxy.}
    \label{fig:r_process_pattern}
\end{figure}

All of the discussed mass loss channels are possible r-process sources. Most advanced are to
date the calculations for the dynamic ejecta, the other channels are still in early exploration stages.
Already the early network calculations for the dynamic ejecta of compact binary mergers\cite{freiburghaus99b} 
showed a very good agreement with the observed solar system abundance pattern, see Fig.~\ref{fig:r_process_pattern},
left panel.
Since no weak interactions were included in these calculations, the electron fraction $Y_e$ was varied in a low
range considered appropriate for neutron stars. These results showed a very good agreement with the solar
system pattern beyond $A\approx 130$ and in particular produced the heaviest elements up to the platinum
peak without any fine tuning of the parameters, but hardly produced yields below $A=130$. Such calculations 
have been more and more refined over recent
years\cite{goriely05,goriely11a,roberts11,korobkin12a,bauswein13a,mendoza_temis14} and they all agree that 
the dynamic ejecta are a very promising site for (at least) the heaviest elements ($A>130$). The abundance pattern has been 
shown\cite{korobkin12a} to be virtually independent of the astrophysical parameters of the merging binary system (masses,
mass ratio, whether the binary system consists of two neutron stars or a neutron star and a black hole) so that every 
CBM produces the same, very robust abundance pattern.
This property is, of course, very interesting from a galactochemical point of view, since such robust abundance patterns are
actually observed for the heaviest elements ($Z\ge56$) in metal-poor stars\cite{sneden08}. Moreover, studies that vary the 
neutron star EOS\cite{bauswein13a} come to
the conclusion that the abundance is also rather stable against such variations. A recent study\cite{mendoza_temis14}
has varied nuclear mass models and investigated the reason for the robustness. The authors suggest
that a requirement for the robustness is that at the moment of freeze-out a much larger mass is in the 
fissioning region (nucleon number $A > 250$) than in the second r-process peak and above ($120 < A < 180$).\\
While the results are --for fixed nuclear physics ingredients-- very robust with respect to a change 
of the astrophysical parameters, they show a substantial sensitivity with respect to the variation of the nuclear 
physics input, they are particularly sensitive  to the distribution of the fission 
products\cite{korobkin12a,eichler14,mendoza_temis14} which have a particular influence on the abundance 
pattern after the second peak. Although the abundance pattern globally agrees 
well with the solar system abundances, a close inspection shows in a number of 
studies\cite{freiburghaus99b,metzger10b,roberts11,korobkin12a,bauswein13a,goriely13,rosswog14a} a slight, 
but systematic shift (with respect to the solar pattern) of the calculated results towards higher $A$. These are 
attributed to ``late'' neutrons, e.g. from $\beta$-delayed neutron emission, but the agreement can be 
improved\cite{eichler14,caballero14} if $\beta$-decay rates are applied that are faster 
than those derived from the frequently used Finite Range Droplet Model\cite{moeller95}. Such increased rates are actually
consistent with recent experimental data for nuclei close to the r-process path\cite{kurtukian14} and recent shell model 
calculations\cite{suzuki12,zhi13}. \\
The nucleosynthesis in the other channels has so far been much less explored. In both the neutrino-driven
wind and the disk-dissolution channel matter stays much longer, roughly a viscous time scale 
Eq.~(\ref{eq:tau_visc})\footnote{In cases where a stable neutron star results the time scales could be as long as
the neutrino cooling time scales of several seconds, see Eq.~(\ref{eq:nu_diff_time_ns}).},
at larger temperatures and in a neutrino background field. Therefore, the weak interactions have time to substantially
increase the electron fractions of the finally unbound material. The first hydrodynamic studies of these 
channels\cite{perego14b,just14} find that they produce less heavy r-process elements and with 
``non-robust'' abundance patterns that depend on the exact details on the merged system. Again, such a behaviour
is consistent with the observation of metal-poor stars\cite{sneden08} that show substantial deviations from the
solar system pattern for $Z<56$ on a star-by-star basis. The wind- and disk-yields complement the dynamic 
ejecta at nucleon numbers below $A \approx 130$ and show broad consistency with the solar abundance pattern.\\
Relativistic gravity also seems to have a substantial influence on the resulting nucleosynthesis. This is mainly because
neutron stars are more compact and the dynamics close to merger is impacted by GR effects. Overall, one expects a more violent
dynamics, larger temperatures in the merged remnant and therefore an enhanced likelyhood to achieve a broader
distribution of $Y_e$ values. A recent study\cite{wanajo14} including general relativity and a detailed neutrino 
treatment reports, for example, on a reproduction of the whole r-process range, however, without discussing
in detail the channels and mechanisms. Closely related to the impact of GR is the equation of state and the
GR-effects can be enhanced by particularly soft equations of state, while stiffer ones yield results that are closer
to those employing Newtonian gravity.

\subsection{Galactic chemical evolution}
\label{sec:chem_evol}
As outlined in Sec.~\ref{sec:nucleo_channels}, all recent studies based on hydrodynamic source plus nuclear network
simulations come to the conclusion that compact binary mergers are excellent candidates for a major r-process
source, at the very least for the heaviest nuclei ($A > 130$), but maybe even for all r-process nuclei. But there are, of 
course, also constraints from galactochemical evolution and the observed elemental scatter in metal-poor stars 
should be different depending on whether the majority of r-process comes from core-collapse supernovae
or from compact binary mergers. For example, as discussed in Sec.~\ref{sec:mass_loss_channels} the typical
ejecta mass of CBM of $\sim 0.01$ \Msun would --together with typical rate estimates-- be able to account for
most/all r-process material. If core-collapse supernovae, in contrast, would be main source and each supernovae
would contribute, the average ejecta mass per event would only be $\sim 10^{-5}$ \msun. Moreover, as will be
discussed in the sGRB context, see Sec.~\ref{sec:GRB}, the large sensitivity of the gravitational wave inspiral times
to the initial binary separation and eccentricity, see Eq.~(\ref{eq:tau_insp}) and Fig.~\ref{fig:inspiral}, naturally
leads to a large spread in delay times after star formation. If endowed with substantial kick velocities, one expects
a large variety of merger sites: some would merge close to their birth places in the galactic plane while others
would merge in the outskirts of galaxies, see the right panel of Fig.~\ref{fig:r_process_pattern} for a sketch.
Such a distribution is indeed observed for short GRBs (also thought to result from CBM), they show
typical projected offsets of $\sim 5$ kpc, see Sec.~\ref{sec:GRB_confrontation}. This also implies a large range
of ambient matter densities encountered for different merger sites and therefore a different ``stopping length'' 
and volume with which the ejecta can mix. Thus, the galactochemical imprint of SN and CBM is likely quite different.\\
Initial inhomogeneous chemical evolution studies\cite{argast04} ruled out neutron star mergers as dominant 
source of r-process, mainly since --due to the low rates-- the r-process enrichment would not be consistent
with the observations at low metallicity and the scatter in [r-process/Fe] would be much too large. This is influenced
on the one hand by the delay time due to GW inspiral, see Eq.~(\ref{eq:tau_insp}) and Fig.~\ref{fig:inspiral}, and
on the other hand by the question whether the neutron star ejecta mix with the chemical products from the 
supernovae which formed the neutron stars in the first place. There are, however, indications that at least some 
of those supernovae that form neutron stars in surviving double neutron star systems are ``non-standard'', they 
may form low-mass, low-kick neutron stars and eject only very small amounts of heavy 
elements\cite{podsiadlowski04,piran05b,stairs06,kitaura06}.
More recent chemical evolution studies\cite{matteucci14a} come to the conclusion that neutron star mergers could well
be a major contributor to the galactic r-process inventory, although probably not the only one. Another chemical 
evolution study\cite{mennekens14a} concludes that, except for the earliest evolutionary phases, CBM could well 
be the major production sites of r-process elements.\\
Most recently, hydrodynamic studies of galactic enrichments have become available. In particular a high resolution
cosmological hydrodynamics simulation called ``Eris'' has been used that is thought to closely represent the evolution
of the Milky Way\cite{shen14a}. The results of this simulation have been used in a post-processing step to follow
the evolution of r-process enrichment by either type II supernovae or compact binary mergers. Unlike in previous studies, 
the authors find that the nucleosynthetic products from compact binary mergers can be incorporated into stars of 
very low metallicity and at early times, even with a minimum delay time of 100 Myr and come to the conclusion that
compact binary mergers could be the dominant source of r-process in the Galaxy. These results are supported by an
independent hydrodynamic study\cite{vandevoort14a} that uses cosmological zoom-in simulations of a Milky Way-mass 
galaxy from the Feedback In Realistic Environments project\cite{hopkins14a}. Also here, the authors conclude that
the results are consistent with ns mergers being the source of most of the r-process nuclei in the Universe. \\
Such an interpretation is also consistent with a recent study\cite{wallner14} of radioactive $^{244}$Pu in 
Earth's deep-see reservoirs. The authors find that the abundances are two orders of magnitude lower than what 
is expected from a continuous enrichment by supernovae. This points to a very rare actinide production site, 
for example a small subset of actinide-producing supernovae or a compact binary merger origin.

\subsection{``Macronovae'': electromagnetic transients powered by radioactive 
                    decays from freshly syntesized heavy elements}
\label{sec:transients}
As outlined above, compact binary mergers eject, via several channels, initially extremely 
neutron-rich matter.  The first and best-studied channel are the dynamic ejecta. Li and 
Paczynski\cite{li98} were the first to realize that if indeed rapid neutron capture occurs
in the ejected material\cite{lattimer74,eichler89,freiburghaus99b} then the radioactivity
in the ejecta should cause an electromagnetic transient following a merger. Such transients
are often referred to as ``kilonovae''\cite{metzger10b} or ``macronovae''\cite{kulkarni05}. 
But since only little mass it ejected ($m_{\rm ej}\sim 0.01$ \msun) at large velocities
($v_{\rm ej} \sim 0.1$ c), the expected light curve evolution is faster and the peak 
luminosity is lower than in the type Ia case. The bulk of the energy release due to 
r-process occurs on a time scale of $\sim$ second \cite{freiburghaus99b,goriely05,metzger10a,roberts11}
when the matter density is still so large that radiation is completely trapped and the 
radioactive heating is used up for accelerating the adiabatic expansion of the ejecta\footnote{For the 
nuclear heating history see, for example, Korobkin et al.\cite{korobkin12a}, their Fig.7.}. It is 
only when the expansion time scale becomes comparable 
to the radiative diffusion time scale that  substantial 
electromagnetic emission occurs\cite{arnett82}. The diffusion time scale and therefore
the time to transparency depend on how opaque the material is with respect to its radiation 
content, but unfortunately the opacities are to date maybe the largest uncertainty in modeling macronovae.\\
For simplicity, the first study of Li and Paczynski  (and all subsequent studies until recently) 
assumed for a lack of better knowledge that the relevant opacities would be of order 
$\kappa \sim 0.1$ cm$^2$/g, similar to the line opacities of iron-group elements\cite{pinto00}.
The composition of the dynamic ejecta, however, is very different from any other known cosmic
explosion and in particular very different from any type of supernova. While the latter produce elements up to 
the iron group near $Z = 26$, the dynamic ejecta of neutron star mergers consist,
due to the extreme initial neutron-richness of the decompressed r-process material,
almost entirely  of r-process elements up to the third peak near $Z \approx 90$, e.g. 
Fig.~\ref{fig:r_process_pattern}. To date, not much is known about the relevant opacities of such material,
but it was recently realized by Kasen and collaborators\cite{kasen13a,barnes13a}
that the opacities have previously been underestimated by orders of magnitude.
The literally millions of lines, Doppler-broadened by the differential velocities of the
ejecta, contribute a pseudo-continuum of bound-bound opacity. Photons trying to
escape the ejecta come into resonance with multiple transitions one by one, resulting in 
a very large opacity. The bulk of the opacity is thereby provided by ions with a particularly complex structure. In 
particular lanthanides ($58 \le Z \le 72$) have --due to the complicated structure 
of their valence f shells-- been found to be major sources of opacity. While detailed
knowledge of opacities is still lacking, the currently existing calculations indicate opacities that
are $\sim$ two orders of magnitude larger than what was previously assumed\cite{barnes13a,kasen13a,tanaka13a}, 
$\kappa \sim 10$ cm$^2$/g, even if only small amounts of lanthanides are present. As a result, 
radiation is much longer trapped and matter becomes transparent only much later, at lower 
temperatures and luminosities and peaking in the near-IR band rather than, as initially found, 
in the optical. Using such increased opacities, recent 
studies\cite{barnes13a,tanaka13a,grossman14a,kasen14} found the resulting 
``macronovae'' should peak after typically a week rather than the original estimate
of half a day.\\
Simple order of magnitude estimates\cite{li98,metzger10b,grossman14a} can be obtained in the following way.
The diffusion time scale for the ejecta with radius $R$ is approximately\footnote{See, e.g., Piran et al. (2013)\cite{piran13a}.}
\be
\tau_{\rm diff} \sim \frac{m_{\rm ej} \kappa}{4 \pi c R}.
\ee
The peak emission is expected\cite{arnett82} when this is equal to the expansion time 
scale $\tau_{\rm exp}=R/v_{\rm ej}$, so at a radius
\be
R_{\rm peak}\sim \sqrt{\frac{m_{\rm ej} \kappa v_{\rm ej}}{4 \pi c}}
= 1.3 \times 10^{15} \; {\rm cm}  \; \left( \frac{m_{\rm ej}}{10^{-2} M_\odot} \right)^{1/2} 
                            \left( \frac{\kappa}{10 \; {\rm cm^2/g}} \right)^{1/2} \left( \frac{v_{\rm ej}}{0.1 c} \right)^{1/2},
\ee
which is reached after
\be
\tau_{\rm peak} \sim \sqrt{\frac{m_{\rm ej} \kappa}{4 \pi c v_{\rm ej}}}
= 4.9 \; {\rm days} \; \left( \frac{m_{\rm ej}}{10^{-2} M_\odot} \right)^{1/2} 
                            \left( \frac{\kappa}{10 \; {\rm cm^2/g}} \right)^{1/2} \left( \frac{0.1 c} {v_{\rm ej}}\right)^{1/2}.
\label{eq:tau_peak}
\ee
If we assume that the late-time radioactive energy release rate can be approximated as a power
 law\cite{li98,metzger10b,rosswog14a}, $\dot{\epsilon}= \dot{\epsilon}_0 (t/t_0)^{-\alpha}$ ($\alpha= 1.3$), 
we can estimate the peak bolometric luminosity  as
\bea
L_{\rm peak}&\sim& m_{\rm ej} \dot{\epsilon}(t_{\rm peak})
= \dot{\epsilon}_0 t_0^\alpha (4 \pi c)^{\frac{\alpha}{2}} \left( \frac{v_{\rm ej}}{\kappa}\right)^{\frac{\alpha}{2}}  
    m_{\rm ej}^{1-\frac{\alpha}{2}}\\
&=& 2.5 \times 10^{40} \; \frac{\rm erg}{\rm s} \; \left( \frac{v_{\rm ej}}{0.1 c}\right)^{\frac{\alpha}{2}}
         \left( \frac{10 \; \rm cm^2/g}{\kappa}\right)^{\frac{\alpha}{2}}  \left( \frac{m_{\rm ej}}{10^{-2} M_\odot}  \right)^{1-\frac{\alpha}{2}},
\label{eq:L_peak}
\eea
where we have used the numerical values as determined in Korobkin et al. (2012). Using the Stefan-Boltzmann law, 
$L= 4 \pi R^2 \sigma_{SB} T^4$, one can obtain a rough estimate for the effective temperature at peak as
\be
T_{\rm peak}^{\rm eff}= 2200 {\rm K} \left( \frac{10 \; \rm cm^2/g}{\kappa}\right)^{\frac{\alpha+2}{8}} 
                                                           \left( \frac{v_{\rm ej}}{0.1 c}\right)^{\frac{\alpha-2}{8}}
                                                           \left( \frac{10^{-2} M_\odot}{m_{\rm ej}}  \right)^{\frac{\alpha}{8}} .
\ee
These simple estimates also illustrate the dramatic effect that the changed opacities have: instead of a transient that peaks in
the blue part of the optical spectrum after $\sim 1$ day (for iron-type opacities), one now expects a peak after $\sim 1$ week
in the near-IR.\\
Early optical searches for macronova emission were carried out for several short 
GRBs\cite{bloom06,berger09,kocevski10,perley09,rowlinson10,perley12}, but none 
of them provided convincing evidence for a macronova. The first serious evidence
came from a search in the near-IR band in the aftermath of GRB 130603B using
a combination of ground-based observations at $<2$ days and HST observations
at about 9 and 30 days after the burst\cite{tanvir13a,berger13b}. The difference
between the observations revealed the presence of a near-IR point source at 9
days after the GRB, this source had disappeared by the time of the second observation.
These observation lead both groups\cite{tanvir13a,berger13b} to conclude  that 
the most natural explanation was a radioactively powered transient due to decaying, 
freshly produced r-process elements, in other words by a macronova.\\
While overall compatible with the expectations for a macronova caused by the extremely 
neutron-rich ejecta, the inferred values\cite{berger13b,tanaka13a,grossman14a} 
for the ejecta masses ($m_{\rm ej}\approx 0.03 - 0.08$ \msun) and velocities\cite{piran14a}
($v_{\rm ej}\approx 0.2c$) are in a plausible, though slightly uncomfortably large regime
for a near-equal mass nsns merger which is considered the most likely event. A 
natural way out would be a CBM with a mass ratio substantially different from unity, either 
nsns or nsbh, which would explain both larger ejecta masses and 
velocities\cite{oechslin07a,rosswog13b}. Such systems, however, are expected to
occur less frequently than $q\approx 1$ nsns-mergers. 
While r-process powered macronova are a natural interpretation, there is still room
for alternative models\cite{kisaka14,takami14b} and further macronova detections are eagerly awaited to
settle the case.

\section{Compact binary mergers as central engines of short gamma-ray bursts}
\label{sec:GRB}
Gamma-ray bursts are intense flashes of soft gamma-rays that are 
detectable approximately once per day and that reach Earth from random directions.
There were early hints \cite{mazets81,norris84,dezalay92}  on a structured 
duration distribution, but this became only firmly established through 
the work of Kouveliotou et al. \cite{kouveliotou93} who found a minimum in the duration 
distribution around $\sim$ 2 s. They also realized that bursts below this time scale 
are consistently harder in their spectra than the bursts above it. This lead to the establishment of two 
classes of bursts, "short, hard bursts" (sGRBs; $\tau \sim 0.3$ s) and "long, soft bursts" 
(lGRBs; $\tau \sim 30$ s)\footnote{This classification has, however, been subject to criticism \cite{zhang07,zhang09}.}.
Their spectra in the conventional $\nu - \nu F_{\nu}$ coordinates are usually
fitted by a broken power law with a pronounced peak around $\sim$ 400 KeV energies
(for sGRBs; $\sim 200$ keV for lGRBs)\cite{paciesas03,ghirlanda09,ghirlanda11}.\\
Compact binary mergers consisting of two neutron stars or a neutron star and a black
hole  have been suggested as one of the first progenitor models  
\cite{blinnikov84,paczynski86,goodman86,goodman87,eichler89,narayan92}, actually
long before the bimodality of the GRB distribution\cite{kouveliotou93}, their cosmological
origin\cite{gehrels05,hjorth05,bloom06} (and therefore the distance and energy scales) and their 
types of host galaxies\cite{prochaska06,fong13,berger14a} were firmly established. The CBM model has survived 
the confrontation with three decades of new observations and many of the arguments that were 
brought forward in the very early papers are still considered valid today.\\
The discussion here will be centered around the role of compact binary mergers, for more exhaustive, 
general discussions of  GRBs and their properties, we refer to the excellent 
reviews\cite{zhang04,piran05a,meszaros06,lyutikov06,lee07,nakar07,gehrels09,kumar14}  that
exist on the topic.

\subsection{Confronting the CBM model with observations}
\label{sec:GRB_confrontation}
We will here summarise to which extent CBMs explain the observed properties of sGRBs
and where tensions between the model and observations exist. Alternative suggestions 
are briefly discussed in Sec.~\ref{sec:alternatives}.\\ 

{\em Energy requirements}\\
Although often paraphrased as "the biggest explosions in the Universe" the energy requirements
of sGRBs are large, but still moderate compared to the rest mass energy of a solar mass,
1 \msun $c^2 \approx 1.8 \times 10^{54}$ erg.
The observed, "isotropic" $\gamma$-energies (i.e. assuming that the radiation is emitted isotropically) 
are in the range of $E_{\gamma, \rm iso}= 10^{48}... 10^{52}$ erg\cite{berger07,nysewander09,berger11}, which 
correspond to "true" energies, if the radiation is emitted into a solid angle with half-opening angle $\theta$, of 
$E_{\gamma}= 10^{46} ... 10^{50} \; {\rm erg} \; (\theta/8^\circ)^2$. In fact, for the small subset of sGRBs
where opening angle information is available the corrected values for both $\gamma-$ray and kinetic energies
indicate $E_{\rm \gamma} \sim E_{\rm kin}\sim 10^{49}$ erg \cite{fong14}.
The main energy reservoir that can be tapped in a CBM is the released gravitational energy which 
is $E_{\rm grav}\sim G M_{\rm tot}^2/R$, where $M_{\rm tot}$ its the total binary mass and $R$ a separation not much larger
than  the Schwarzschild radius $R_s= 2G M_{\rm tot}/c^2= 9 \; {\rm km} \; (M_{\rm tot}/3 M_\odot)$. So that, in 
principle, an energy approaching $\sim M_{\rm tot}c^2/2$, corresponding to several times $10^{53}$
erg, is available. Therefore, the energy requirements are not a serious challenge, even if only 
moderate efficiencies should be applicable for transforming released energy into $\gamma$-rays.\\

{\em Variability time scale}\\
The energy output of sGRBs can vary substantially on a time scale of a few ms\cite{berger11,berger14a}. 
This is the time scale that is naturally expected from a compact binary system. 
Both from the dynamical time scale of a neutron star, see Eq.~(\ref{eq:tau_dyn_ns}), and from
the orbital period at the innermost, stable circular orbit (ISCO) of a stellar-mass black hole,
see Eq.~(\ref{eq:tau_dyn_disk}), one would naturally expect variations of order milliseconds.\\

{\em Duration}\\
It is commonly expected that an accretion disk is needed to transform the released gravitational binding energy 
into ultra-relativistic outflow and finally into the observed radiation. 
If one assumes that such extremely rapidly accreting systems can be described by the simple thick-disk 
estimate Eq.~(\ref{eq:tau_visc}), there is good agreement with the typical
duration of sGRBs\cite{berger14a}, $\tau_{\rm sGRB} \sim 0.3$ s.\\

{\em Host galaxies/environment}\\
The host galaxies of short bursts are systematically different from those of long bursts. While the latter occur
in unusually bright star-forming regions \cite{fruchter06,svensson10}, the former occur in both early- ($\sim 20\%$)
 and late-type ($\sim 80\%$) galaxies\cite{gehrels09,fong13}. This points to a generally older stellar population and a substantial spread
of ages. This is actually expected for compact binary mergers and has in fact been a prediction of the merger model\cite{fryer99a}. 
As discussed in Sec.~\ref{sec:time_scales}, the inspiral time due to the emission of gravitational waves is a
sensitive function of the initial orbital period $P$ and the eccentricity $e$, see Eq.~(\ref{eq:tau_insp}) and 
Fig.~\ref{fig:inspiral}. Since the initial values of both quantities are set by evolutionary processes 
of the stellar binary progenitor, one expects indeed a broad spread of inspiral times. Due to kicks 
that the neutron stars receive at birth large initial eccentricities 
may be common.\footnote{Note, however, that some systems 
may have formed via different mechanisms. In the case of the double pulsar system, PSR J0737-3039A/B, the 
low space velocity, the comparatively low mass of pulsar B, the low orbital eccentricity ($e=0.09$) and the 
location only $\sim 50$ pc from the Galactic midplane are all consistent with pulsar B having been formed by 
a mechanism {\em different} from the usually assumed core-collapse of a helium star\cite{piran05b,stairs06}. In 
fact, progenitor masses possibly substantially below 2\Msun are kinematically favored.} If the binary as a whole 
receives systemic kick, substantial distances can be travelled during the inspiral time and one expects a broad 
distribution of merger offsets with respect to their host galaxies. The distributions obtained from binary evolution 
calculations\cite{bloom99,fryer99a,rosswog03c,belczynski06} agree remarkably well with those observed for 
sGRBs \cite{fong10,church11,fong13}. The observed projected offsets range from 0.5 to 75 kpc with a median 
value of $\approx$ 5 kpc. These results are interpreted as projected kick velocities\cite{fong13} from 
$v_{\rm kick}\sim 20$ to 140 km s$^{-1}$ (a median value of $\sim 60$ km s$^{-1}$) and consistent with the 
binary population synthesis results ($v_{\rm kick}\sim 5$ to 500 km s$^{-1}$) for nsns mergers 
\cite{fryer97,fryer98,wang06,wong10}. Also, the sGRB redshifts expected from CBM models are in agreement
with those observed\cite{berger11,fong13,berger14a} ($\sim 0.1 < z < 1.5$).\\

{\em Beaming and event rates}\\
The angular structure of the outflow can be inferred from so-called achromatic ``jet-breaks''\cite{sari99,rhoads99,panaitescu05}: 
the light curve decay begins to steepen roughly simultaneously in all wavelength bands. This is caused by
a combination of ``relativistic beaming''\cite{meszaros99,panaitescu99,rhoads99,sari99} and, in addition,
by the sideway expansion of the jetted outflow\cite{rhoads99,sari99,kumar03,granot12}.\\
If the outflow of sGRBs is collimated\footnote{We refer to ``collimated'' to describe the geometry of the
outflow and to ``beamed'' for the special-relativistic aberration effect.} into a half-opening angle $\Theta$, the emission from each emitting patch
is ``beamed'' into a forward cone\cite{rybicki79} with half-opening angle  
$1/\gamma$, where $\gamma$ is the local Lorentz factor. As a consequence, only observers that
happen to be in the beam cone can observe the emission and for them it looks like a spherical flow with
the properties of the local patch. While the jet spreads sideways and the outflow becomes decelerated by the interaction with 
the ambient medium, the relativistic beaming angle increases, and once similar to the true geometric collimation 
angle, $1/\gamma \sim \Theta$, the jetted nature of the outflow becomes noticeable and a jet break in the
lightcurve indicates that so far only a small patch had been visible.
If redshift, kinetic energy and ambient matter density are known, one can infer the geometric jet opening 
$\theta$ from the time of the jet break\footnote{See, for example,
Eq.(33) in Nakar (2007)\cite{nakar07}.}. Obviously, the value of this angle has severe implications for both the overall
energy budget, see above, and the true event rate: for small opening angles, most bursts are unobservable and
the true rate is larger by a ``beaming factor'' $f_b = 4\pi/\Delta \Omega \approx 2/\Theta^2$, where
$\Delta \Omega$ is the solid angle of the beam.\\
So far, it has been difficult to observe jet breaks for sGRBs, mainly because of their weak afterglows. 
There is only a handful of cases for jet breaks and they have values of $\sim 10^\circ$ \cite{fong13,berger14a}, 
in rough agreement with the theoretically expected values\cite{rosswog03b,aloy05,nagakura14}. This would imply that only 
roughly one out of 70 sGRBs is detectable and it would translate\cite{fong13} into a true sGRB rate of 
$\approx 20$ yr$^{-1}$ within 200 Mpc, roughly the distance accessible to Advanced LIGO/VIRGO.\\

{\em "Macronova" emission}\\
``Macronovae'', see Sec.~\ref{sec:transients}, have actually been a prediction of the compact binary 
merger model. The detection of the first the macronova event in the aftermath of 
GRB130603B\cite{tanvir13a,berger13b}  is broadly consistent with the expectations for radioactively 
powered transients that result from the extremely neutron-rich dynamic
ejecta that produce very heavy nuclei ($A>130$). If the current ideas about other mass loss channels, see 
Sec.~\ref{sec:mass_loss_channels}, are correct, there should be additional transients with different properties, 
also powered by radioactivity.\\

{\em Lorentz factors}\\
GRBs, both long and short ones, involve highly relativistic bulk flows. This was suggested as a solution of the so-called
``compactness problem''\cite{ruderman75,schmidt78} and later also confirmed by direct observations\cite{frail97}. 
The clue to the compactness problem is
to understand, how a luminous source can vary on time scales of order milliseconds and still emit optically thin radiation.
Assume a (slowly moving) sphere of radius $R$ that emits radiation and changes substantially on a time scale $\delta t$.
If it would ``switch off'' immediately, it would take at least the light travel time difference between closest and the farthest visible
point of the sphere, $R/c$, to conveigh this information to an observer. Turned around, (assuming a non-relativistic source)
an observer would conclude that the source must be of a size $R < c \delta t \simeq 300 \; {\rm km} (\delta t$/1 ms).  Now 
inserting typical numbers of GRBs, such a size would imply enormous optical depths ($\tau \sim 10^{13}$) for photons with 
respect to the at photon energies of $\sim$ MeV copiously present electron-positron pairs\cite{schmidt78}, in stark contrast 
to the observed, optically thin radiation\footnote{This argument had originally been used to place an upper distance limit on GRB sources.}.\\
If the source is instead moving relativistically to the observer, the source can be larger by a factor of $\sim \gamma^2$ 
(because the observer can, due to relativistic beaming, only see a small patch of the source, but the real source size is
much larger) and also the photon energy in the emission frame is lower than the observed one. The requirement $\tau<1$ 
can then be used to place lower limits on the Lorenz factor, estimates typically favour $10^2 < \gamma < 10^3$, see e.g.
Lithwick \& Sari \cite{lithwick01} for a detailed analysis.\\
A second line of argument comes from the onset of the afterglow. Once the blast wave has deposited most of its
energy in an ambient medium, it assumes a self-similar profile\cite{blandford76} and X-ray and optical emissions
decay as powerlaws. During this powerlaw phase the evolution of the Lorentz factor (only weakly depending on the
kinetic energy of the blast and the ambient medium density) is known, and the time of the onset of this powerlaw behaviour
provides a lower limit on the Lorentz factor.\\
The presence of such large Lorentz-factors, is actually a very non-trivial constraint on GRB models. To accelerate to 
an assymptotic Lorentz factor of $\gamma$, a blast of energy $E$, cannot contain more than 
\be
m_{\rm bar}= \frac{E}{\gamma c^2}= 5.5 \times 10^{-9} M_\odot \; \left( \frac{E_{48}} {\gamma_{100}}\right),
\label{eq:baryonic_pollution}
\ee
where we refer to a quantity $X_{n}$ as $X/10^n$ in cgs-units.  It is at least not obvious, how 
a small fraction of the mass can receive such a disproportionate share of the released energy.
How are the deposition of mass and energy so cleanly separated?\\

{\em "Late-time activity''}\\
There are also a number events that show ``late-time activity'' on a time scale much longer than what is naively expected
from the compact binary merger model, see Sec.~\ref{sec:time_scales}. To date this activity is still only incompletely understood.\\
A subset of $\sim 20$ \% of sGRBs\cite{norris10,berger14a} shows after the first, ``prompt'' spike somewhat softer, extended
$\gamma$-ray emission that lasts for $\sim 10$ - 100 s and is sometimes delayed in its onset. This emission component
was first seen in a stacking analysis of short BATSE bursts \cite{lazzati01} and subsequently observed in a number of other bursts
\cite{connaughton02,frederiks04,villasenor05,barthelmy05,perley09}. In some cases, the fluence in this ``extended emission''
can exceed the one in the prompt spike, in the extreme example of GRB080503 the extended emission continues up to
$\sim 200$ s and dominates the fluence of the initial spike by a factor of 32\cite{perley09}.\\
Another type of late-time activity are  X-ray flares following the prompt $\gamma$-ray emission which have actually been
observed for both long an short bursts\cite{burrows05,nousek06,chincarini07,falcone07,margutti10,chincarini11,margutti11}.
In some cases, such flares have been seen many hours after the burst, e.g. GRB050724 showed a significant X-ray flare
at $\sim 14$ hours after the burst, for GRB 130603B excess X-ray emission has been detected more than a day after the
main burst.  For sGRBs such flares occur for different types of GRB host galaxies and both for cases with extended emission 
and without\cite{margutti11}.\\
Such time scales that exceed both the dynamical and the viscous time scales by many orders of magnitude 
are clearly uncomfortable for the compact binary merger model and had not been expected
prior to detection. Given that main power source of GRBs is accretion it is natural to start from the hypothesis that it
may also power this late-time activity. The dissipation parameter $\alpha$ in Eq.(\ref{eq:tau_visc}) is not well known,
but it seems unlikely to have a value that maintains accretion for much longer than seconds. Compact binary mergers 
(in particular those with mass ratios deviating from unity) also eject a few percent of a solar mass into nearly unbound
orbits, and this material while providing an energy reservoir $> 10^{50}$ erg provides a time scale, $\tau_{fb}$, that can
easily exceed both the dynamical and the viscous time scale by orders of magnitude \cite{rosswog07a,lee07,faber06b}.
However, at least in its simplest form, it may be challenged to release large amounts of energy at very late times 
($\gg$ seconds). Dynamical collisions between compact objects, see Sec.~\ref{sec:dyn_coll}, carry some promise in that 
respect\cite{lee07,east12a,east12b,rosswog13a} since each of the several close encounters produces a tidal tail
that serves as a mass reservoir far from the engine. Most likely, however, they are too rare to explain the common
late-time activity phenomena. While the initial, purely ballistic 
fallback ideas may have been too simple to explain the observations, it has become clearer over the last decade
that compact binary mergers eject mass via several channels, see Sec.~\ref{sec:heavy_elements}, some driven
by neutrino- and/or nuclear processes, and the interaction of these different mass loss channels may yield
a much more complicated picture than just the ballistic fallback. In that sense, there is some promise that also the 
late-time activity can be reconciled with compact binary mergers, but this question is certainly far from being 
understood at the moment.

\subsection{How to launch ultra-relativistic outflow}
Models for the ultra-relativistic outflow must --apart from questions of collimation and stability-- first
and foremost explain how some fraction of matter can acquire such a disproportionate share of
the energy. The very large energy to rest mass ratio required to produce the inferred ultra-relativistic outflow, see 
Eq.~(\ref{eq:baryonic_pollution}), is a very non-trivial requirement and to date it is still a matter of debate
how this is achieved in GRB. One of the mechanisms 
suggested\cite{eichler89,narayan91,mochkovitch93,ruffert97a,popham99,rosswog02b,rosswog03b,fryer03,birkl07,zalamea11}  
to achieve this is the annihilation of neutrino anti-neutrino  pairs, 
$\bar{\nu}_i \nu_i \rightarrow e^- e^+$. The annihilation cross-section is very small, but the neutrino luminosities
are huge, see Sec.~\ref{sec:neutrino}, and if a small fraction of the neutrino energy can be deposited in a baryon free region
the typical energy of a short GRB, $\sim 10^{48}$ erg, see Sec.~\ref{sec:GRB_confrontation}, could plausibly be 
reached. Due to the dependence of the annihilation rate on a factor $\mu \equiv 1 - \cos{\theta}$, where $\theta$ is the
collision angle between neutrino and anti-neutrino\footnote{In the explicit expression\cite{goodman87,cooperstein87a}, 
there are actually two such terms, one is proportional to $\mu$, the other to $\mu^2$.}, headon collisions are
favoured. Therefore, spacetime curvature or thick disk geometries can enhance the annihilation rate. 
A necessary requirement for neutrinos from an accretion disk is that they can escape on short enough time
scale so that they are not just advected into the bh. But as estimated in Sec.~\ref{sec:neutrino} and further detailed 
by numerical simulations, the neutrino escape time scales are substantially shorter than the dynamic disk time scales.\\
Large neutrino
luminosities, however, can also lead to strong neutrino-driven winds that can make it impossible to reach the required
energy to rest mass ratio. This effect is particularly pronounced for remnants that contain a central 
HMNS\cite{dessart09,perego14b}. The situation is less critical for black hole disk systems, 
for which numerical simulations\cite{fernandez13b,just14} find much weaker neutrino-driven
winds. A recent study\cite{murguia14} came to the conclusion that the HMNS must collapse within 
$\sim 100$ ms, otherwise baryonic pollution precludes the emergence of ultra-relativistic outflow.\\
Magnetic mechanisms are another broad class of plausible mechanisms. They are, for example, required to extract the
energy of a spinning black hole via the Blandford-Znajek mechanism\cite{blandford77}, or they can more directly
be transformed in outflow by buoyancy instabilities with subsequent reconnection events\cite{narayan92,kluzniak98a,price06}
or as some form of extreme pulsar\cite{usov92,duncan92,rosswog03c,dai06} where the rotational energy is tapped. All these processes 
require very strong, near-equipartition magnetic field strengths to be effective. This is, however, likely to be achieved, 
at least for the short relevant time scale, since neutron stars are from the beginning endowed with strong magnetic 
fields and the merger process offers ample possiblities to amplify 
them\cite{price06,anderson08b,rezzolla11,zrake13,giacomazzo14,kiuchi14}.\\
Of course, it is certainly a possibility that the diversity of observed bursts is related to
several of these mechanisms work in concert.

\subsection{Double neutron star vs neutron star black hole binaries}                                                                                                       \label{sec:nsn_vs_nsbh}   
  \begin{figure}[htb]
    \vspace*{-2cm}
    \centerline{\includegraphics[width=9cm,angle=-90]{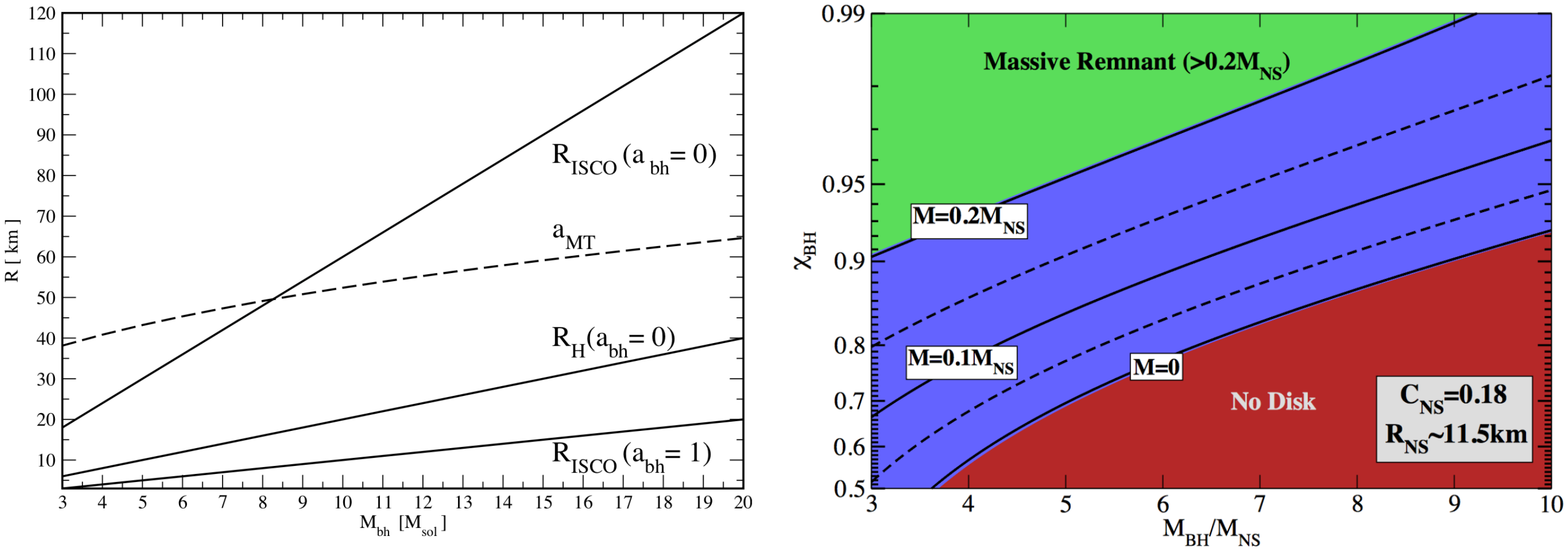}}
    \vspace*{-2cm}
    \caption{Left panel: illustration of the mass transfer separation $a_{\rm MT}$ with respect to horizon 
                  and innermost stable circular orbits based on Newtonian estimates (for fiducial 
                  neutron star properties, $M_{\mathrm{ns}}= 1.4$ \Msun and $R_{\mathrm{ns}}= 12$ km).
                  Right panel: results of an analytical model\cite{foucart12} fitted to numerical simulations 
                  for the mass that remains at late times outside the hole. The remaining mass depends apart
                  the black hole mass $M_{\rm bh}$ and spin $\chi_{\rm BH}$ (= $a_{\rm bh}$ in our notation)
                  also on the neutron star compactness $C_{\rm NS}$. Results for more/less compact neutron
                 stars can be found in the original paper\cite{foucart12}.}
    \label{fig:BH_radii}
\end{figure}
Traditionally, nsns and nsbh systems have been considered as ``standard GRB'' progenitor,
often without much distinction between them, since they were thought to both lead to 
the most likely engine, a stellar mass bh surrounded by an accretion disk. In recent years,
however, this picture has become more differentiated. On the one hand it has become clear
that bh formation in the nsns case does not have to happen early on and may actually in
some cases not happen at all, see the discussion in Sec.~\ref{sec:time_scales}. The presence 
of a HMNS, in turn, implies strong neutrino-driven winds (Sec.~\ref{sec:nu_winds}), which, 
in turn, pose a potential threat to the ability of launching ultra-relativistic outflow. The nsbh
case supposedly produces a baryon-cleaner environment (mainly due to the absence of
a HMNS) and it is more likely to show imprints from jet precession\cite{stone13}.
But not every combination of bh masses and spins is actually able to form a substantial
accretion torus around the hole. This is mainly because the separation where the neutron star
starts transferring mass/is disrupted grows with a lower power of the bh mass $M_{\rm bh}$ than 
the innermost stable circular orbit (ISCO). Therefore for large enough
bh masses the neutron star is swallowed before being disrupted so that no torus can form.\\
The final answer  requires 3D numerical simulations with the relevant physics, 
but a qualitative idea can still be gained from simple estimates. Mass transfer is expected to set 
in when the Roche volume becomes comparable to the volume of the neutron star. By applying 
Paczynski's estimate for the Roche lobe radius\cite{paczynski71} and equating it with the ns 
radius, one finds that the onset of mass transfer (which we use here as a proxy for the tidal 
disruption radius) can be expected near a separation of
\be
a_{\rm MT}= 2.17 R_{\rm ns} \left( \frac{1 + q}{q} \right)^{1/3}
\approx 26 \; {\rm km} \left( \frac{R_{\rm ns}}  {12 \; {\rm km}} \right)
\left( \frac{1 + q}{q} \right)^{1/3}   .
\ee
Since $a_{\rm MT}$ grows, in the limit where the mass ratio $q\equiv M_{\rm ns}/M_{\rm bh} \ll 1$, 
only proportional to $M_{\rm bh}^{1/3}$, but the ISCO and the event horizon grow $\propto M_{\rm bh}$, 
the onset of mass transfer/disruption can  take place inside the ISCO for large bh masses. At the very high end of bh masses, the 
neutron star is swallowed as whole without being disrupted at all. A qualitative illustration 
(for fiducial neutron star properties, $M_{\rm ns}= 1.4$ \Msun and $R_{\rm ns}= 12$ km) is 
shown in Fig.~\ref{fig:BH_radii}, left panel. Roughly, already for black holes near $M_{\rm bh}\approx 8$ 
\Msun the mass transfer/disruption occurs near the ISCO which makes it potentially difficult 
to form a massive torus from ns debris. So, low mass black holes are clearly preferred as
GRB engines. Numerical simulations \cite{faber06b} have shown, however,  that even if the 
disruption occurs deep inside the ISCO this does not necessarily mean that all the matter is
doomed to fall straight into the hole and  a torus can still possibly form.\\
\begin{figure}[pb]
\centerline{\psfig{file=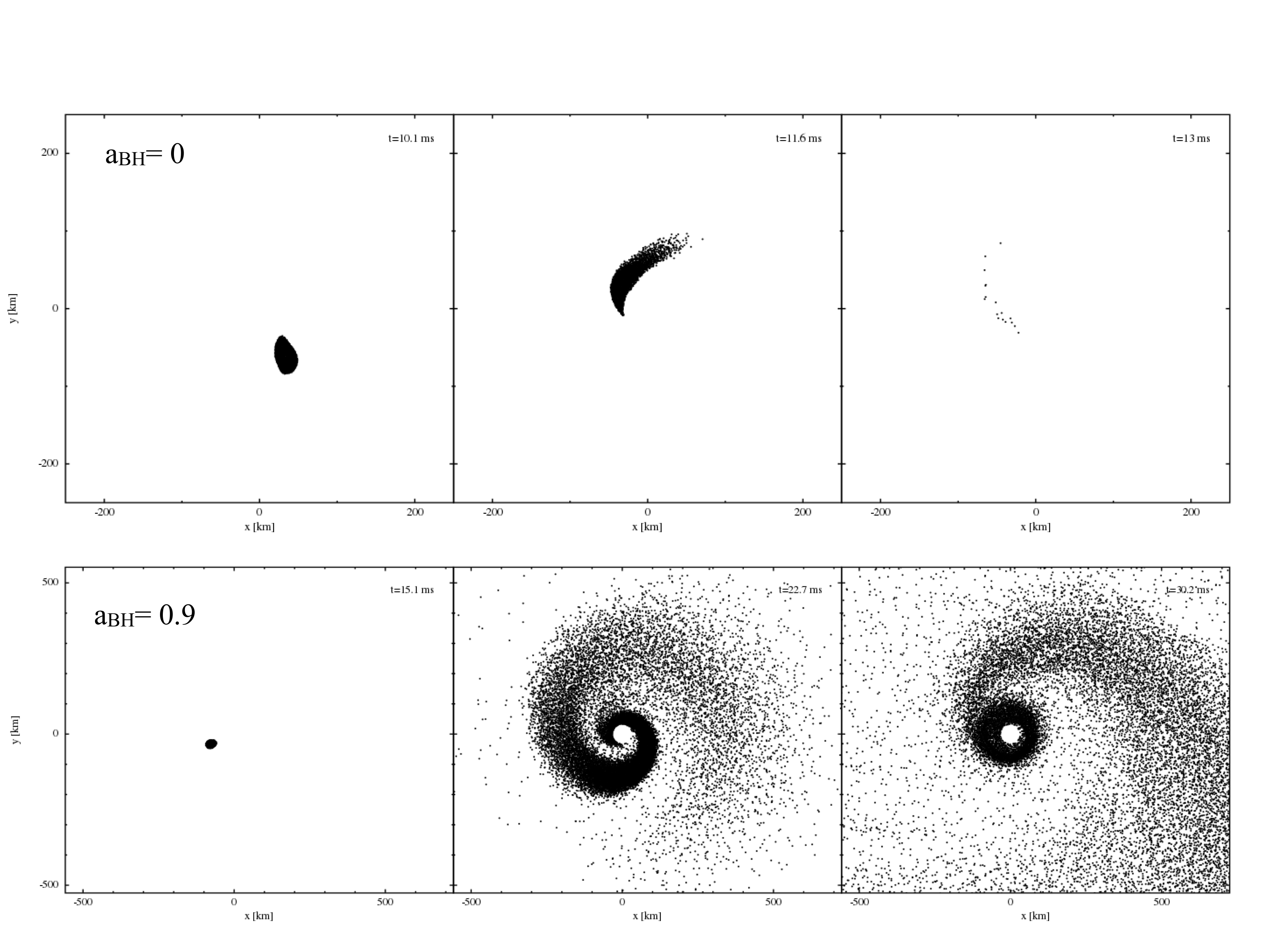,width=12.5cm,angle=0}}
\vspace*{8pt}
\caption{Impact of black hole spin on disk formation. The panels shown in the upper row
refer to the merger of a neutron star with a Schwarzschild bh ($a_{\rm bh}= 0$), while the 
lower row shows the merger with a rapidly spinning Kerr bh ($a_{\rm bh}= 0.9$). In both cases 
a 1.3 \Msun neutron star is set on a circular orbit with $r_c= 8 G M_{\rm bh}/c^2$. Due to the ISCO lying at a
larger radius, for the $a_{\rm bh}=0$ case the black is swallowed almost completely (only very few 
SPH particles are left at $t=13$ ms), while for the highly spinning case a massive accretion disk
forms (last panel, second row). Note that the panels in the two rows show different scales.}
\label{fig:spin_comparison}
\end{figure}
The bh spin also plays a crucial role for the question of disk formation, since it determines
the location of the ISCO, for maximally spinning Kerr bhs, for example, it is located at only 
$R_{\rm ISCO}= G M_{\rm bh}/c^2$. This, of course, has a serious impact on the dynamics and in 
particular on the resulting tori. This is illustrated in Fig.~\ref{fig:spin_comparison} where
two nsbh binary systems are simulated (using the SPH method) that are identical (initially circular radii
with $r_c= 8 G M_{\rm bh}/c^2$ of a 1.3 \Msun ns around a spatially fixed Kerr bh) 
up to the bh spin parameter. In the first case of a non-rotating bh (upper row, $a_{bh}= 0$) the ns is swallowed
completely within $\sim 13$ ms, while for the rapidly spinning case ($a_{bh}= 0.9$) a massive
torus forms.\\
Fully, relativistic simulations are computationally very expensive, but it is possible
to devise simple analytical models\cite{foucart12} for the mass remaining outside the 
bh\footnote{This refers not only to disk mass, but also contains in addition dynamic ejecta and
tidal tail.} that have a physically intuitive functional form and whose parameters can be fitted 
to the results of relativistic nsbh simulations. Once fitted, the results depend on the black hole
mass $M_{\rm bh}$, its dimensionless spin parameter $a_{\rm BH}$ and the neutron star compactness
$C_{\rm ns}$. The result for the probably most likely neutron star compactness is shown
in Fig.~\ref{fig:BH_radii}, right panel (their quantity $\chi_{\rm BH}$ is what we refer to as $a_{\rm bh}$). 
The results for more extreme compactness values can be found in the original paper\cite{foucart12}. 
According to this model, a 10 \Msun bh needs to have at least a spin of $a_{\rm bh}\approx 0.9$ to
form a disk in the disruption of a ``standard'' neutron star, for larger/smaller stars (13.5 km/9.5 km) 
spin values of 0.84/0.98 are needed. \\ 
When discussing disk formation in a GRB context, it is worth keeping in mind that even 
seemingly small disk masses allow, at least in principle, for the extraction of energies,
\be
E_{\rm extr} \sim 1.8\times 10^{51} \; {\rm erg} \; \left( \frac{\epsilon}{0.1}\right) \left( \frac{M_{\rm disk}}{0.01 M_\odot}\right),
\ee
that are large enough to accommodate the isotropic gamma-ray energies, $E_{\gamma, \rm iso} \sim 10^{50}$ erg, that have been inferred
for short bursts \cite{berger14a}. Collimation into half-opening angle $\Theta$ further reduces the
requirement by $\Theta^2/2$.\\
If the bh mass distribution in nsbh binary systems is indeed peaked\cite{oezel10,belczynski08} around $8 - 10$ \Msun
then the large spins required to form a disk ($a_{\rm bh} \geq 0.8$) may seriously constrain the parameter space.
Recent spin measurements in X-ray binaries\cite{mcclintock14} find in four out of ten cases spins $>0.8$, thus 
giving some hope that such high spins may possibly be realized also in nsbh binaries.
A recent study\cite{stone13} finds that for most of their Monte Carlo models between 10 and 30\% of the nsbh
mergers could be able to launch a sGRB, but the fraction could essentially vanish for bottom-heavy bh spin
distributions. One should therefore be prepared that the contribution of nsbh mergers to the observed sGRB rate 
could actually  be small.

\subsection{Gravitational wave-driven binary mergers vs. dynamic collisions}
\label{sec:dyn_coll}
Traditionally, mostly gravitational wave-driven binary systems such as the Hulse-Taylor pulsar
\cite{taylor89,weisberg10} have been discussed as GRB engines, dynamical collisions among neutron stars
and black holes in stellar systems with large number densities such as Globular Clusters
have been considered a very unlikely event\cite{ruffert98}. More recently, however, such collisions 
have received a fair amount of attention\cite{kocsis06a,oleary09,lee10a,east12a,east12b,kocsis12,gold12,
gold13,rosswog13a,moldenhauer14}. Unfortunately, their rates are at least as difficult to estimate as those of 
GW-driven mergers.\\
Collisions differ from gravitational wave-driven mergers in a number of ways. For example, since gravitational wave emission
of eccentric binaries efficiently removes angular momentum in comparison to energy, primordial binaries will have radiated away
their eccentricity and will finally merge from a nearly circular orbit. On the contrary, binaries that have formed dynamically,
say in a globular cluster, start from a small orbital separation, but with large eccentricities and may not have had the time
to circularize until merger. This leads to pronouncedly different gravitational wave signatures, ``chirping'' signals of
increasing frequency and amplitude for mergers and initially well-separated, repeated GW bursts that continue from
minutes to days in the case of collisions. Moreover,
compact binaries are strongly gravitationally bound at the onset of the dynamical merger phase while collisions
have total orbital energies close to zero and need to shed energy and angular momentum via GW emission
and/or through mass shedding episodes in order to form a single remnant. Due to the strong dependence on the
 impact parameter and the lack of strong constraints on it, one therefore expects a much larger variety of dynamical
behavior for collisions than for mergers.\\
A study that discussed various aspects of such collisions in detail\cite{lee10a}  came to the conclusion
that such encounters could possibly produce an interesting contribution to the observed GRB rate. Indeed,
such collisions typically yield large accretion tori masses, peak neutrino luminosities\cite{rosswog13a} in 
excess of $10^{53}$ erg/s, but in some cases substantially more, and ample possibilities to amplify initial seed
magnetic fields\cite{price06,anderson08b,obergaulinger10,rezzolla11,zrake13,giacomazzo14,kiuchi14}. Perhaps most 
interestingly, since they typically suffer several close encounters before
they can merge into a single object, and since each close encounter results in tidal tail, they produce a more 
complicated environment of the central merger remnant. This stores in particular subtantial amounts of mass
far from the remnant which only falls back on longer time scales and is maybe related to the late-time activity 
of some sGRBs, see Sec.~\ref{sec:GRB_confrontation}.\\ 
Collisions, however, also share a possible caveat with the merger cases: at least in the nsns case they likely 
produce very strong neutrino-driven winds that could prevent triggering a short GRB. More recently, it was concluded
that collisions cannot be very frequent, since they eject very large amounts of neutron-rich matter. The dominant
nsbh collision channel\footnote{Lee et al.\cite{lee10a} find that nsbh collisions dominate by approximately a factor 
of five over nsns collisions.} ejects typically 0.15 \Msun per event\cite{lee10a,rosswog13a},  which --from cosmic
nucleosynthesis arguments-- restricts the (nsns + nsbh) collision rate to a maximum of $\approx 10$\% of the
estimated nsns merger rate\cite{rosswog13a}.  The rare nsbh collision cases, however, would result in particularly
bright ``macronova'' emission with
\be
L_{\rm peak} \approx 10^{41} {\rm erg/s} \; \left( \frac{v_{\rm ej}}{0.2 c}\right)^{\frac{\alpha}{2}}
         \left( \frac{10 \; \rm cm^2/g}{\kappa}\right)^{\frac{\alpha}{2}}  \left( \frac{m_{\rm ej}}{0.15 M_\odot}  
         \right)^{1-\frac{\alpha}{2}}, 
\ee
peaking after approximately 13 days, see Eqs.~(\ref{eq:tau_peak}) and (\ref{eq:L_peak}).

\subsection{Alternative models}
\label{sec:alternatives}
As an alternative to the common bh-plus-disk systems, magnetars have been
suggested as central engines of sGRBs, both for long and for short 
bursts\cite{usov92,duncan92,metzger08b,rowlinson10,bucciantini12,rowlinson13,metzger14b}. 
Magnetar-like objects could be formed under a variety of circumstances. 
Giant flares from ``young'', extragalactic magnetars with long recurrence times 
could plausibly be responsible for a fraction of the events observed as 
sGRBs\cite{abbott08,ofek08,hurley10,abadie12a}. 
A progenitor population of young magnetars, however, would be coincident with
star forming regions, therefore they can be excluded as the source of the majority of
sGRBs. ``Old'' magnetar-like objects could also be formed through the accretion 
induced collapse (AIC) of an accreting WD or, maybe, as a result of a double white 
dwarf merger\cite{metzger08b,bucciantini12}. For such old magnetars, however, no
strong kicks are expected and therefore their distribution with respect to their host
galaxy would differ from what is observed for sGRBs. It had also been 
suspected\cite{shibata00,rosswog03c,dai06} that nsns mergers may in some cases
result in a rapidly spinning, high-field magnetar-like object and the recent limits
on the maximum mass of neutron stars $>2$ \Msun\cite{demorest10,antoniadis13} 
only make this a more intriguing possibility, see Sec.~\ref{sec:time_scales}.
A magnetar engine would have a number of obvious advantages. There would be  
natural formation mechanisms and a long-lived or stable central object could plausibly
be responsible for the observed late-time activity. A magnetar formed as a result of
a nsns-merger would obviously inherit all the benefits of the nsns merger engine model. 
In addition, having a magnetar at the engine of long and short bursts would naturally 
explain similarities between both types of bursts and with an energy emission rate 
(in the framework of the magnetic dipole model) of
\be
\dot{E}_{\rm md} \propto B^2 R^6 \omega^4, 
\ee
with $B$, $R$  and $\omega$ being magnetic field strength, radius  and angular frequency,
such an engine could naturally produce the large variety of sGRBs that is observed.\\
One could argue, though, that such a model, that depends on twelve powers of poorly
known quantities has not much predictive power. Maybe more severe, there would be some
challenge to explain both the prompt emission and the late-time activity, since the presence
of a magnetar would initially produce an enormous baryon-loading which could prevent a 
burst to form in the first place. In fact, a recent study\cite{murguia14}  finds that collapse 
must occur before $\sim 100$ ms, otherwise a sGRB would be prevented. Maybe the choking of
a GRB can be circumvented if the burst is only launched after a neutrino cooling time of several 
seconds when the neutrino-driven wind has ceased\cite{bucciantini12}.\\
Another alternative engine has been suggested\cite{macfadyen05} as a response to the discovery 
of late-time X-ray flares for short GRBs. In this model, the idea is that a neutron star accretes from a
non-degenerate companion until it collapses into a bh surrounded by an accretion disk, generally
considered the standard engine of GRBs. The X-ray flares would result when part of the relativistic
ejecta interact with the extended companion star. In such a model, the neutron star would need to accrete 
$\sim 0.7$ \msun, not an entirely trivial task given that the Eddington accretion rate is 
$\dot{M}_{\rm Edd} \sim 10^{-8}$ \msun/yr. Another question that would need further exploration is
whether/for which EOS and rotation rate combination of the collapsing neutron star a substantial disk 
can form outside the ISCO of the forming black hole.\\
In summary, there is likely room for several sGRB engines and maybe this would explain in 
part their diversity. However, while none of the suggestions is completely free of open questions,
it is probably a fair statement that the best model to date for the {\em bulk} of sGRBs are compact 
binary mergers with --among them-- a slight preference for nsns mergers, see the discussion in 
Sec.~\ref{sec:nsn_vs_nsbh}.

\section{Summary}
\label{sec:summary}

The last years have seen a tremendous progress in our understanding of compact binary mergers,
both on the theoretical and the observational side, and we have provided here an overview over
various of their facets.\\
The last decade has in particular witnessed the first detection of short GRB afterglows and subsequent observations
have provided a wealth of information about the environments in which short GRBs occur. Most of the observed properties
find a natural explanation in the compact binary merger model, but some properties such as very late
activity keep posing a problem and are not understood. They could either point to the merger event being
way more complicated than imagined in today's models or, alternatively, to engines that are different from 
the standard black hole plus disk picture.\\
By now, practically all theoretical models agree that compact binary mergers eject enough material to be at least
a major source of the heavy ($A>130$) r-process and nucleosynthesis calculations show good agreement with
the observed abundance distributions in this regime. It has also become clear that a merger has several ways to 
enrich its host galaxy with neutron-rich matter: in addition to ``dynamic ejecta''  there are also neutrino- and/or 
magnetically driven winds and accretion tori that become unbound on viscous times scales and each of these channels 
has likely different properties. There are even recent indications that the combination of the different channels 
may actually not just produce the heaviest elements, but even the whole r-process range, although contributions
from other sources such as a supernovae are plausible, in particular for the lighter r-process elements.
A question where no consensus exists yet, is whether compact binary mergers as dominant r-process source 
would be consistent with the chemical evolution of galaxies. While earlier work excluded them as dominant sources,
more recent studies based on hydrodynamic simulations come to the conclusion that compact binary mergers
may well be consistent with the elemental scatter that is observed in stars of different ages.\\
Maybe one of the most exiting new developments is the recent discovery of a credible radioactively powered,
electromagnetic transient candidate, also known as "macronova" or "kilonova", that has been observed
in the aftermath of a short GRB. In particular its time scale of about one week and the peak in the nIR are consistent
with having been produced by very heavy (and therefore opaque) r-process material.
If this is the correct interpretation, it connects for the first time directly short GRBs with compact binary mergers
and  r-process nucleosynthesis. Such transients will also increase the science returns in the era of gravitational 
wave astronomy that hopefully soon will begin.

\section*{Acknowledgments}
This work has been supported by the Deutsche Forschungsgemeinschaft (DFG) 
under grant number RO-3399/5-1 and by the Swedish Research Council (VR) 
under grant 621-2012-4870. It is a pleasure to thank Almudena Arcones, Oleg Korobkin, 
Tsvi Piran, Enrico Ramirez-Ruiz and Friedrich-Karl Thielemann for stimulating discussions.
Some of the results discussed in this article have been obtained on the facilities of the 
The North-German Supercomputing Alliance (HLRN).
\bibliographystyle{ws-ijmpd}
\bibliography{astro_SKR_spec}

\end{document}

%% file: own_latex_commands_i.tex
\def\msun{M$_{\odot}$}
\def\Msun{M$_{\odot}$ }
\def\be{\begin{equation}}
\def\ee{\end{equation}}
\def\bi{\begin{itemize}}

\def\ei{\end{itemize}}
\def\ben{\begin{enumerate}}
\def\een{\end{enumerate}}
\def\bea{\begin{eqnarray}}
\def\eea{\end{eqnarray}}
\def\bt{\begin{tabbing}}
\def\et{\end{tabbing}}
\def\gcc{gcm$^{-3}$}
\def\Gcc{gcm$^{-3} \;$}
\def\edo{\end{document}}

\def\M4{M$_4$}
\def\M6{M$_6$}
\def\lcm6{LCM$_6$}
\def\qcm6{QCM$_6$}

\def\wh3{W$_{\rm H,3}$}
\def\wh4{W$_{\rm H,4}$}
\def\wh5{W$_{\rm H,5}$}
\def\wh6{W$_{\rm H,6}$}
\def\wh7{W$_{\rm H,7}$}

%% file: ms.bbl
\hyphenation{Post-Script Sprin-ger}
\begin{thebibliography}{100}

\bibitem{hulse75}
R.~A. {Hulse} and J.~H. {Taylor}, {\em ApJL} {\bf 195} (January 1975) L51.

\bibitem{weisberg10}
J.~M. {Weisberg}, D.~J. {Nice} and J.~H. {Taylor}, {\em ApJ} {\bf 722} (October
  2010) 1030.

\bibitem{maggiore08}
M.~Maggiore, {\em Gravitational Waves} (Oxford University Press, Oxford, 2008).

\bibitem{einstein15}
A.~Einstein, {\em Preuss. Akad. Wiss. Berlin} {\bf Sitzber. 47}  (1915)   831.

\bibitem{dyson20}
F.~Dyson, A.~Eddington and C.~Davidson, {\em Phil. Trans. Roy. Soc.} {\bf 220A}
   (1920)   291.

\bibitem{froeschle97}
M.~{Froeschle}, F.~{Mignard} and F.~{Arenou}, { {Determination of the PPN
  Parameter gamma with the HIPPARCOS Data}}, in {\em Hipparcos - Venice '97\/},
   eds. R.~M. {Bonnet}, E.~{H{\o}g}, P.~L. {Bernacca}, L.~{Emiliani},
  A.~{Blaauw}, C.~{Turon}, J.~{Kovalevsky}, L.~{Lindegren}, H.~{Hassan},
  M.~{Bouffard}, B.~{Strim}, D.~{Heger}, M.~A.~C. {Perryman} and L.~{Woltjer},
  ESA Special Publication, Vol.~402 (August 1997), pp. 49--52.

\bibitem{shapiro64}
I.~I. {Shapiro}, {\em Physical Review Letters} {\bf 13} (December 1964) 789.

\bibitem{ciufolini04}
I.~{Ciufolini} and E.~C. {Pavlis}, {\em Nature} {\bf 431} (October 2004) 958.

\bibitem{everitt11}
C.~W.~F. {Everitt}, D.~B. {Debra}, B.~W. {Parkinson}, J.~P. {Turneaure}, J.~W.
  {Conklin}, M.~I. {Heifetz}, G.~M. {Keiser}, A.~S. {Silbergleit}, T.~{Holmes},
  J.~{Kolodziejczak}, M.~{Al-Meshari}, J.~C. {Mester}, B.~{Muhlfelder}, V.~G.
  {Solomonik}, K.~{Stahl}, P.~W. {Worden}, Jr., W.~{Bencze}, S.~{Buchman},
  B.~{Clarke}, A.~{Al-Jadaan}, H.~{Al-Jibreen}, J.~{Li}, J.~A. {Lipa}, J.~M.
  {Lockhart}, B.~{Al-Suwaidan}, M.~{Taber} and S.~{Wang}, {\em Physical Review
  Letters} {\bf 106} (June 2011)   221101.

\bibitem{einstein37}
A.~{Einstein} and N.~{Rosen}, {\em Journal of The Franklin Institute} {\bf 223}
  (January 1937) 43.

\bibitem{bondi59}
H.~{Bondi}, F.~A.~E. {Pirani} and I.~{Robinson}, {\em Royal Society of London
  Proceedings Series A} {\bf 251} (June 1959) 519.

\bibitem{kramer08}
M.~{Kramer} and I.~H. {Stairs}, {\em Annual Review of Astronomy and
  Astrophysics} {\bf 46} (September 2008) 541.

\bibitem{kramer09b}
M.~{Kramer} and N.~{Wex}, {\em Classical and Quantum Gravity} {\bf 26} (April
  2009)   073001.

\bibitem{burgay03}
M.~{Burgay}, N.~{D'Amico}, A.~{Possenti}, R.~N. {Manchester}, A.~G. {Lyne},
  B.~C. {Joshi}, M.~A. {McLaughlin}, M.~{Kramer}, J.~M. {Sarkissian},
  F.~{Camilo}, V.~{Kalogera}, C.~{Kim} and D.~R. {Lorimer}, {\em Nature} {\bf
  426} (December 2003) 531.

\bibitem{lyne04}
A.~G. {Lyne}, M.~{Burgay}, M.~{Kramer}, A.~{Possenti}, R.~N. {Manchester},
  F.~{Camilo}, M.~A. {McLaughlin}, D.~R. {Lorimer}, N.~{D'Amico}, B.~C.
  {Joshi}, J.~{Reynolds} and P.~C.~C. {Freire}, {\em Science} {\bf 303}
  (February 2004) 1153.

\bibitem{kramer13}
M.~{Kramer}, { {Probing gravitation with pulsars}}, in {\em IAU Symposium\/},
  ed. J.~{van Leeuwen}, IAU Symposium, Vol.~291 (March 2013), pp. 19--26.

\bibitem{manchester05}
R.~N. {Manchester}, G.~B. {Hobbs}, A.~{Teoh} and M.~{Hobbs}, {\em VizieR Online
  Data Catalog} {\bf 7245} (August 2005)  ~0.

\bibitem{lorimer08}
D.~R. {Lorimer}, {\em Living Reviews in Relativity} {\bf 11} (November 2008)
  ~8.

\bibitem{damour09}
T.~{Damour}, { {Binary Systems as Test-Beds of Gravity Theories}}, in {\em
  Astrophysics and Space Science Library\/},  eds. M.~{Colpi}, P.~{Casella},
  V.~{Gorini}, U.~{Moschella} and A.~{Possenti}, Astrophysics and Space Science
  Library, Vol.~359 (2009), pp. 1--4020.

\bibitem{wex14}
N.~{Wex}, {\em arXiv:1402.5594}  (February 2014).

\bibitem{kramer14}
M.~{Kramer}, {\em International Journal of Modern Physics D} {\bf 23} (December
  2014)   30004.

\bibitem{fryer99a}
C.~L. {Fryer}, S.~E. {Woosley} and D.~H. {Hartmann}, {\em ApJ} {\bf 526}
  (November 1999) 152.

\bibitem{dominik12}
M.~{Dominik}, K.~{Belczynski}, C.~{Fryer}, D.~E. {Holz}, E.~{Berti},
  T.~{Bulik}, I.~{Mandel} and R.~{O'Shaughnessy}, {\em ApJ} {\bf 759} (November
  2012)  ~52.

\bibitem{dominik13}
M.~{Dominik}, K.~{Belczynski}, C.~{Fryer}, D.~E. {Holz}, E.~{Berti},
  T.~{Bulik}, I.~{Mandel} and R.~{O'Shaughnessy}, {\em Ap} {\bf 779} (December
  2013)  ~72.

\bibitem{dominik14}
M.~{Dominik}, E.~{Berti}, R.~{O'Shaughnessy}, I.~{Mandel}, K.~{Belczynski},
  C.~{Fryer}, D.~{Holz}, T.~{Bulik} and F.~{Pannarale}, {\em arXiv:1405.7016}
  (May 2014).

\bibitem{abbott09a}
B.~P. {Abbott}, R.~{Abbott}, R.~{Adhikari}, P.~{Ajith}, B.~{Allen}, G.~{Allen},
  R.~S. {Amin}, S.~B. {Anderson}, W.~G. {Anderson}, M.~A. {Arain} and et~al.,
  {\em Reports on Progress in Physics} {\bf 72} (July 2009)   076901.

\bibitem{harry10}
G.~M. {Harry} and {LIGO Scientific Collaboration}, {\em Classical and Quantum
  Gravity} {\bf 27} (April 2010)   084006.

\bibitem{ligo}
LIGO, {\em http://www.ligo.caltech.edu} .

\bibitem{sengupta10}
A.~S. {Sengupta}, {LIGO Scientific Collaboration} and {Virgo Collaboration},
  {\em Journal of Physics Conference Series} {\bf 228} (May 2010)   012002.

\bibitem{virgo}
VIRGO, {\em http://www.virgo.infn.it} .

\bibitem{somiya12}
K.~{Somiya}, {\em Classical and Quantum Gravity} {\bf 29} (June 2012)   124007.

\bibitem{aVIRGO15}
F.~{Acernese}, M.~{Agathos}, K.~{Agatsuma}, D.~{Aisa}, N.~{Allemandou},
  A.~{Allocca}, J.~{Amarni}, P.~{Astone}, G.~{Balestri}, G.~{Ballardin} and
  et~al., {\em Classical and Quantum Gravity} {\bf 32} (January 2015)   024001.

\bibitem{blanchet06}
L.~{Blanchet}, {\em Living Reviews in Relativity} {\bf 9} (June 2006)  ~4.

\bibitem{futamase07}
T.~{Futamase} and Y.~{Itoh}, {\em Living Reviews in Relativity} {\bf 10} (March
  2007)  ~2.

\bibitem{shibata11}
M.~{Shibata} and K.~{Taniguchi}, {\em Living Reviews in Relativity} {\bf 14}
  (August 2011)  ~6.

\bibitem{faber12}
J.~A. {Faber} and F.~A. {Rasio}, {\em Living Reviews in Relativity} {\bf 15}
  (July 2012)  ~8.

\bibitem{rosswog14c}
S.~{Rosswog}, {\em arXiv:1406.4224}  (June 2014).

\bibitem{lattimer74}
J.~M. Lattimer and D.~N. Schramm, {\em ApJ, (Letters)} {\bf 192}  (1974)
  L145.

\bibitem{lattimer76}
J.~M. Lattimer and D.~N. Schramm, {\em ApJ} {\bf 210}  (1976)   549.

\bibitem{lattimer77}
J.~M. {Lattimer}, F.~{Mackie}, D.~G. {Ravenhall} and D.~N. {Schramm}, {\em ApJ}
  {\bf 213} (April 1977) 225.

\bibitem{peters63}
P.~C. {Peters} and J.~{Mathews}, {\em Physical Review} {\bf 131} (July 1963)
  435.

\bibitem{bildsten92}
L.~{Bildsten} and C.~{Cutler}, {\em ApJ} {\bf 400} (November 1992) 175.

\bibitem{kochanek92}
C.~S. {Kochanek}, {\em ApJ} {\bf 398} (October 1992) 234.

\bibitem{baumgarte00}
T.~W. {Baumgarte}, S.~L. {Shapiro} and M.~{Shibata}, {\em ApJL} {\bf 528}
  (January 2000) L29.

\bibitem{kaplan14}
J.~D. {Kaplan}, C.~D. {Ott}, E.~P. {O'Connor}, K.~{Kiuchi}, L.~{Roberts} and
  M.~{Duez}, {\em ApJ} {\bf 790} (July 2014)  ~19.

\bibitem{kastaun14}
W.~{Kastaun} and F.~{Galeazzi}, {\em arXiv:1411.7975}  (November 2014).

\bibitem{takami14}
K.~{Takami}, L.~{Rezzolla} and L.~{Baiotti}, {\em arXiv:1412.3240}  (December
  2014).

\bibitem{price06}
D.~Price and S.~Rosswog, {\em Science} {\bf 312}  (2006)   719.

\bibitem{ostriker68}
J.~P. {Ostriker} and P.~{Bodenheimer}, {\em ApJ} {\bf 151} (March 1968)   1089.

\bibitem{shibata06c}
M.~{Shibata} and K.~{Taniguchi}, {\em Phys. Rev. D} {\bf 73} (March 2006)
  064027.

\bibitem{hotokezaka11}
K.~{Hotokezaka}, K.~{Kyutoku}, H.~{Okawa}, M.~{Shibata} and K.~{Kiuchi}, {\em
  Phys. Rev. D} {\bf 83} (June 2011)   124008.

\bibitem{hotokezaka13b}
K.~{Hotokezaka}, K.~{Kiuchi}, K.~{Kyutoku}, T.~{Muranushi}, Y.-i. {Sekiguchi},
  M.~{Shibata} and K.~{Taniguchi}, {\em Phys. Rev. D} {\bf 88} (August 2013)
  044026.

\bibitem{demorest10}
P.~B. {Demorest}, T.~{Pennucci}, S.~M. {Ransom}, M.~S.~E. {Roberts} and
  J.~W.~T. {Hessels}, {\em Nature} {\bf 467} (October 2010) 1081.

\bibitem{antoniadis13}
J.~{Antoniadis}, P.~C.~C. {Freire}, N.~{Wex}, T.~M. {Tauris}, J.~P.~W.
  {Verbiest} and D.~G. {Whelan}, {\em Science} {\bf 340} (April 2013)   448.

\bibitem{lattimer12a}
J.~M. {Lattimer}, {\em Annual Review of Nuclear and Particle Science} {\bf 62}
  (November 2012) 485.

\bibitem{yu13}
Y.-W. {Yu}, B.~{Zhang} and H.~{Gao}, {\em ApJL} {\bf 776} (October 2013)   L40.

\bibitem{takami14c}
H.~{Takami}, K.~{Kyutoku} and K.~{Ioka}, {\em Phys. Rev. D} {\bf 89} (March
  2014)   063006.

\bibitem{metzger14b}
B.~D. {Metzger} and A.~L. {Piro}, {\em MNRAS} {\bf 439} (April 2014) 3916.

\bibitem{shakura73}
N.~I. {Shakura} and R.~A. {Sunyaev}, {\em A \& A} {\bf 24}  (1973) 337.

\bibitem{arnett89}
W.~D. {Arnett}, J.~N. {Bahcall}, R.~P. {Kirshner} and S.~E. {Woosley}, {\em
  Annual Review of Astronomy and Astrophysics} {\bf 27}  (1989) 629.

\bibitem{waxman04a}
E.~{Waxman}, {\em New Journal of Physics} {\bf 6} (October 2004)   140.

\bibitem{waxman04b}
E.~{Waxman}, {\em ApJ} {\bf 606} (May 2004) 988.

\bibitem{dermer05}
C.~D. {Dermer} and J.~M. {Holmes}, {\em ApJl} {\bf 628} (July 2005) L21.

\bibitem{waxman06a}
E.~{Waxman}, {\em Nuclear Physics B Proceedings Supplements} {\bf 151} (January
  2006) 46.

\bibitem{waxman97}
E.~{Waxman} and J.~{Bahcall}, {\em Physical Review Letters} {\bf 78} (March
  1997) 2292.

\bibitem{rachen98}
J.~P. {Rachen} and P.~{M{\'e}sz{\'a}ros}, {\em \prd} {\bf 58} (December 1998)
  123005.

\bibitem{waxman99}
E.~{Waxman} and J.~{Bahcall}, {\em Phys. Rev. D} {\bf 59} (January 1999)
  023002.

\bibitem{dermer03}
C.~D. {Dermer} and A.~{Atoyan}, {\em Physical Review Letters} {\bf 91} (August
  2003)   071102.

\bibitem{derishev99}
E.~V. {Derishev}, V.~V. {Kocharovsky} and V.~V. {Kocharovsky}, {\em ApJ} {\bf
  521} (August 1999) 640.

\bibitem{bahcall00}
J.~N. {Bahcall} and P.~{M{\'e}sz{\'a}ros}, {\em Physical Review Letters} {\bf
  85} (August 2000)   1362.

\bibitem{waxman00}
E.~{Waxman} and J.~N. {Bahcall}, {\em ApJ} {\bf 541} (October 2000) 707.

\bibitem{dermer02}
C.~D. {Dermer}, {\em ApJ} {\bf 574} (July 2002) 65.

\bibitem{li02}
Z.~{Li}, Z.~G. {Dai} and T.~{Lu}, {\em A \& A} {\bf 396} (December 2002) 303.

\bibitem{lai94c}
D.~{Lai}, {\em MNRAS} {\bf 270} (October 1994)   611.

\bibitem{glendenning00}
N.~K. {Glendenning}, {\em {Compact Stars}} (2nd. ed., Springer-Verlag, New
  York, Berlin, Heidelberg: Springer, 2000., 2000).

\bibitem{ruffert97a}
M.~{Ruffert}, H.~{Janka}, K.~{Takahashi} and G.~{Schaefer}, {\em A \& A} {\bf
  319} (March 1997) 122.

\bibitem{rosswog03a}
S.~{Rosswog} and M.~{Liebend{\"o}rfer}, {\em MNRAS} {\bf 342} (July 2003) 673.

\bibitem{dessart09}
L.~{Dessart}, C.~D. {Ott}, A.~{Burrows}, S.~{Rosswog} and E.~{Livne}, {\em ApJ}
  {\bf 690} (January 2009) 1681.

\bibitem{perego14b}
A.~{Perego}, S.~{Rosswog}, R.~M. {Cabez{\'o}n}, O.~{Korobkin},
  R.~{K{\"a}ppeli}, A.~{Arcones} and M.~{Liebend{\"o}rfer}, {\em MNRAS} {\bf
  443} (October 2014) 3134.

\bibitem{tubbs75}
D.~Tubbs and D.~Schramm, {\em ApJ} {\bf 201}  (1975)   467.

\bibitem{sekiguchi10b}
Y.~{Sekiguchi}, {\em Classical and Quantum Gravity} {\bf 27} (June 2010)
  114107.

\bibitem{deaton13a}
M.~B. {Deaton}, M.~D. {Duez}, F.~{Foucart}, E.~{O'Connor}, C.~D. {Ott}, L.~E.
  {Kidder}, C.~D. {Muhlberger}, M.~A. {Scheel} and B.~{Szilagyi}, {\em ApJ}
  {\bf 776} (October 2013)  ~47.

\bibitem{neilsen14}
D.~{Neilsen}, S.~L. {Liebling}, M.~{Anderson}, L.~{Lehner}, E.~{O'Connor} and
  C.~{Palenzuela}, {\em Phys. Rev. D} {\bf 89} (May 2014)   104029.

\bibitem{bruenn85}
S.~W. Bruenn, {\em ApJS} {\bf 58}  (1985).

\bibitem{ruffert01}
M.~{Ruffert} and H.-T. {Janka}, {\em A\&A} {\bf 380} (December 2001) 544.

\bibitem{sekiguchi11}
Y.~{Sekiguchi}, K.~{Kiuchi}, K.~{Kyutoku} and M.~{Shibata}, {\em Physical
  Review Letters} {\bf 107} (July 2011)   051102.

\bibitem{rosswog13a}
S.~{Rosswog}, T.~{Piran} and E.~{Nakar}, {\em MNRAS} {\bf 430} (April 2013)
  2585.

\bibitem{foucart14}
F.~{Foucart}, M.~B. {Deaton}, M.~D. {Duez}, E.~{O'Connor}, C.~D. {Ott},
  R.~{Haas}, L.~E. {Kidder}, H.~P. {Pfeiffer}, M.~A. {Scheel} and
  B.~{Szilagyi}, {\em Phys. Rev. D} {\bf 90} (July 2014)   024026.

\bibitem{oezel10}
F.~{{\"O}zel}, D.~{Psaltis}, R.~{Narayan} and J.~E. {McClintock}, {\em ApJ}
  {\bf 725} (December 2010) 1918.

\bibitem{belczynski08}
K.~{Belczynski}, R.~E. {Taam}, E.~{Rantsiou} and M.~{van der Sluys}, {\em ApJ}
  {\bf 682} (July 2008) 474.

\bibitem{burbidge57}
E.~M. {Burbidge}, G.~R. {Burbidge}, W.~A. {Fowler} and F.~{Hoyle}, {\em Reviews
  of Modern Physics} {\bf 29}  (1957) 547.

\bibitem{cameron57b}
A.~G.~W. {Cameron}, {\em Publications of the Astronomical Society of the
  Pacific} {\bf 69} (June 1957)   201.

\bibitem{cowan91}
J.~J. Cowan, F.-K. Thielemann and J.~W. Truran, {\em Phys. Rep.} {\bf 208}
  (1991)   267.

\bibitem{woosley92}
S.~E. Woosley and R.~D. Hoffman, {\em ApJ} {\bf 395}  (1992)   202.

\bibitem{arnould07}
M.~{Arnould}, S.~{Goriely} and K.~{Takahashi}, {\em Phys. Reports} {\bf 450}
  (September 2007) 97.

\bibitem{duncan86}
R.~C. {Duncan}, S.~L. {Shapiro} and I.~{Wasserman}, {\em ApJ} {\bf 309}
  (October 1986) 141.

\bibitem{qian96b}
Y.-Z. {Qian} and S.~E. {Woosley}, {\em ApJ} {\bf 471} (November 1996)   331.

\bibitem{hoffman97}
R.~D. Hoffman, S.~E. Woosley and Y.-Z. Qian, {\em ApJ} {\bf 482}  (1997)   951.

\bibitem{freiburghaus99a}
C.~Freiburghaus, J.~Rembges, T.~Rauscher, E.~Kolbe, F.-K. Thielemann, K.-L.
  Kratz and J.~Cowan, {\em ApJ} {\bf 516}  (1999)   381.

\bibitem{otsuki00}
K.~{Otsuki}, H.~{Tagoshi}, T.~{Kajino} and S.-y. {Wanajo}, {\em ApJ} {\bf 533}
  (April 2000) 424.

\bibitem{thompson01}
T.~A. {Thompson}, A.~{Burrows} and B.~S. {Meyer}, {\em ApJS} {\bf 562}
  (December 2001) 887.

\bibitem{farouqi10}
K.~{Farouqi}, K.-L. {Kratz}, B.~{Pfeiffer}, T.~{Rauscher}, F.-K. {Thielemann}
  and J.~W. {Truran}, {\em ApJ} {\bf 712} (April 2010) 1359.

\bibitem{arcones07}
A.~{Arcones}, H.-T. {Janka} and L.~{Scheck}, {\em A \& A} {\bf 467} (June 2007)
  1227.

\bibitem{fischer10}
T.~{Fischer}, S.~C. {Whitehouse}, A.~{Mezzacappa}, F.-K. {Thielemann} and
  M.~{Liebend{\"o}rfer}, {\em A \& A} {\bf 517} (July 2010)   A80.

\bibitem{huedepohl10}
L.~{H{\"u}depohl}, B.~{M{\"u}ller}, H.-T. {Janka}, A.~{Marek} and G.~G.
  {Raffelt}, {\em Physical Review Letters} {\bf 104} (June 2010)   251101.

\bibitem{roberts10}
L.~F. {Roberts}, S.~E. {Woosley} and R.~D. {Hoffman}, {\em ApJ} {\bf 722}
  (April 2010)   954.

\bibitem{arcones13a}
A.~{Arcones} and F.-K. {Thielemann}, {\em Journal of Physics G Nuclear Physics}
  {\bf 40} (January 2013)   013201.

\bibitem{winteler12b}
C.~{Winteler}, R.~{K{\"a}ppeli}, A.~{Perego}, A.~{Arcones}, N.~{Vasset},
  N.~{Nishimura}, M.~{Liebend{\"o}rfer} and F.-K. {Thielemann}, {\em ApJL} {\bf
  750} (May 2012)   L22.

\bibitem{eichler89}
D.~Eichler, M.~Livio, T.~Piran and D.~N. Schramm, {\em Nature} {\bf 340}
  (1989)   126.

\bibitem{rosswog99}
S.~Rosswog, M.~Liebend\"orfer, F.-K. Thielemann, M.~Davies, W.~Benz and
  T.~Piran, {\em A \&\ A} {\bf 341}  (1999) 499.

\bibitem{freiburghaus99b}
C.~Freiburghaus, S.~Rosswog and F.-K. Thielemann, {\em ApJ} {\bf 525}  (1999)
  L121.

\bibitem{oechslin07a}
R.~{Oechslin}, H.~{Janka} and A.~{Marek}, {\em A \& A} {\bf 467} (May 2007)
  395.

\bibitem{roberts11}
L.~F. {Roberts}, D.~{Kasen}, W.~H. {Lee} and E.~{Ramirez-Ruiz}, {\em ApJL} {\bf
  736} (July 2011)   L21.

\bibitem{goriely11a}
S.~{Goriely}, A.~{Bauswein} and H.-T. {Janka}, {\em ApJL} {\bf 738} (September
  2011)   L32.

\bibitem{korobkin12a}
O.~{Korobkin}, S.~{Rosswog}, A.~{Arcones} and C.~{Winteler}, {\em MNRAS} {\bf
  426} (November 2012) 1940.

\bibitem{wanajo14}
S.~{Wanajo}, Y.~{Sekiguchi}, N.~{Nishimura}, K.~{Kiuchi}, K.~{Kyutoku} and
  M.~{Shibata}, {\em ApJL} {\bf 789} (July 2014)   L39.

\bibitem{just14}
O.~{Just}, A.~{Bauswein}, R.~{Ardevol Pulpillo}, S.~{Goriely} and H.-T. {Janka}
  (June 2014), {\em arXiv:1406.2687}.

\bibitem{rosswog13b}
S.~{Rosswog}, {\em Royal Society of London Philosophical Transactions Series A}
  {\bf 371} (April 2013)   20272.

\bibitem{metzger08a}
B.~D. {Metzger}, A.~L. {Piro} and E.~{Quataert}, {\em MNRAS} {\bf 390} (October
  2008) 781.

\bibitem{beloborodov08}
A.~M. {Beloborodov}, { {Hyper-accreting black holes}}, in {\em American
  Institute of Physics Conference Series\/},  ed. {M.~Axelsson}, American
  Institute of Physics Conference Series, Vol.~1054 (September 2008), pp.
  51--70.

\bibitem{lee09}
W.~H. {Lee}, E.~{Ramirez-Ruiz} and D.~{L{\'o}pez-C{\'a}mara}, {\em ApJL} {\bf
  699} (July 2009) L93.

\bibitem{fernandez13a}
R.~{Fernandez} and B.~D. {Metzger}, {\em ApJ} {\bf 763} (February 2013)   108.

\bibitem{fernandez13b}
R.~{Fernandez} and B.~D. {Metzger}, {\em MNRAS} {\bf 435} (October 2013) 502.

\bibitem{rosswog02b}
S.~{Rosswog} and E.~{Ramirez-Ruiz}, {\em MNRAS} {\bf 336} (October 2002) L7.

\bibitem{rosswog03b}
S.~{Rosswog} and E.~{Ramirez-Ruiz}, {\em MNRAS} {\bf 343} (August 2003) L36.

\bibitem{rosswog03c}
S.~{Rosswog}, E.~{Ramirez-Ruiz} and M.~B. {Davies}, {\em MNRAS} {\bf 345}
  (November 2003) 1077.

\bibitem{shibata11b}
M.~{Shibata}, Y.~{Suwa}, K.~{Kiuchi} and K.~{Ioka}, {\em ApJL} {\bf 734} (June
  2011)   L36.

\bibitem{kiuchi12b}
K.~{Kiuchi}, K.~{Kyutoku} and M.~{Shibata}, {\em Phys. Rev. D} {\bf 86}
  (September 2012)   064008.

\bibitem{siegel14a}
D.~M. {Siegel}, R.~{Ciolfi} and L.~{Rezzolla}, {\em ApJL} {\bf 785} (April
  2014)  ~L6.

\bibitem{kiuchi14}
K.~{Kiuchi}, K.~{Kyutoku}, Y.~{Sekiguchi}, M.~{Shibata} and T.~{Wada}, {\em
  Phys. Rev. D} {\bf 90} (August 2014)   041502.

\bibitem{kyutoku14}
K.~{Kyutoku}, K.~{Ioka} and M.~{Shibata}, {\em MNRAS} {\bf 437} (January 2014)
  L6.

\bibitem{metzger15}
B.~D. {Metzger}, A.~{Bauswein}, S.~{Goriely} and D.~{Kasen}, {\em MNRAS} {\bf
  446} (January 2015) 1115.

\bibitem{kaeppeler89}
F.~K\"appeler, H.~Beer and K.~Wisshak, {\em Rep. Prog. Phys.} {\bf 52}  (1989)
   945.

\bibitem{sneden08}
C.~{Sneden}, J.~J. {Cowan} and R.~{Gallino}, {\em Annual Review of Astronomy
  and Astrophysics} {\bf 46} (September 2008) 241.

\bibitem{mcmillan11a}
P.~J. {McMillan}, {\em MNRAS} {\bf 414} (July 2011) 2446.

\bibitem{abadie10}
J.~{Abadie}, B.~P. {Abbott}, R.~{Abbott}, M.~{Abernathy}, T.~{Accadia},
  F.~{Acernese}, C.~{Adams}, R.~{Adhikari}, P.~{Ajith}, B.~{Allen} and et~al.,
  {\em Classical and Quantum Gravity} {\bf 27} (September 2010)   173001.

\bibitem{bauswein14b}
A.~{Bauswein}, R.~{Ardevol Pulpillo}, H.-T. {Janka} and S.~{Goriely}, {\em
  ApJL} {\bf 795} (November 2014)  ~L9.

\bibitem{bauswein13a}
A.~{Bauswein}, S.~{Goriely} and H.-T. {Janka}, {\em ApJ} {\bf 773} (August
  2013)  ~78.

\bibitem{foucart13}
F.~{Foucart}, M.~B. {Deaton}, M.~D. {Duez}, L.~E. {Kidder}, I.~{MacDonald},
  C.~D. {Ott}, H.~P. {Pfeiffer}, M.~A. {Scheel}, B.~{Szilagyi} and S.~A.
  {Teukolsky}, {\em Phys. Rev. D} {\bf 87} (April 2013)   084006.

\bibitem{hotokezaka13a}
K.~{Hotokezaka}, K.~{Kiuchi}, K.~{Kyutoku}, H.~{Okawa}, Y.-i. {Sekiguchi},
  M.~{Shibata} and K.~{Taniguchi}, {\em Phys. Rev. D} {\bf 87} (January 2013)
  024001.

\bibitem{kyutoku13}
K.~{Kyutoku}, K.~{Ioka} and M.~{Shibata}, {\em Phys. Rev. D} {\bf 88} (August
  2013)   041503.

\bibitem{rosswog05a}
S.~{Rosswog}, {\em ApJ} {\bf 634} (December 2005) 1202.

\bibitem{shen98a}
H.~Shen, H.~Toki, K.~Oyamatsu and K.~Sumiyoshi, {\em Nuclear Physics} {\bf A
  637}  (1998)   435.

\bibitem{shen98b}
H.~Shen, H.~Toki, K.~Oyamatsu and K.~Sumiyoshi, {\em Progress of Theoretical
  Physics} {\bf 100}  (1998) 1013.

\bibitem{metzger14a}
B.~D. {Metzger} and R.~{Fern{\'a}ndez}, {\em MNRAS} {\bf 441} (July 2014) 3444.

\bibitem{kiuchi12}
K.~{Kiuchi}, Y.~{Sekiguchi}, K.~{Kyutoku} and M.~{Shibata}, {\em Classical and
  Quantum Gravity} {\bf 29} (June 2012)   124003.

\bibitem{mclaughlin05}
G.~C. {McLaughlin} and R.~{Surman}, {\em Nuclear Physics A} {\bf 758} (July
  2005) 189.

\bibitem{surman06}
R.~{Surman}, G.~C. {McLaughlin} and W.~R. {Hix}, {\em ApJ} {\bf 643} (June
  2006) 1057.

\bibitem{surman08}
R.~{Surman}, G.~C. {McLaughlin}, M.~{Ruffert}, H.~{Janka} and W.~R. {Hix}, {\em
  ApJL} {\bf 679} (June 2008) L117.

\bibitem{wanajo12}
S.~{Wanajo} and H.-T. {Janka}, {\em ApJ} {\bf 746} (February 2012)   180.

\bibitem{caballero12}
O.~L. {Caballero}, G.~C. {McLaughlin} and R.~{Surman}, {\em ApJ} {\bf 745}
  (February 2012)   170.

\bibitem{perego14a}
A.~{Perego}, E.~{Gafton}, R.~{Cabez{\'o}n}, S.~{Rosswog} and
  M.~{Liebend{\"o}rfer}, {\em A \& A} {\bf 568} (August 2014)   A11.

\bibitem{kaeppeli11}
R.~{K{\"a}ppeli}, S.~C. {Whitehouse}, S.~{Scheidegger}, U.-L. {Pen} and
  M.~{Liebend{\"o}rfer}, {\em ApJS} {\bf 195} (August 2011)  ~20.

\bibitem{goriely05}
S.~{Goriely}, {\em Nuclear Physics A} {\bf 752} (April 2005) 560.

\bibitem{mendoza_temis14}
J.~{de Jesus Mendoza-Temis}, G.~{Martinez-Pinedo}, K.~{Langanke}, A.~{Bauswein}
  and H.-T. {Janka}, {\em arXiv:1409.6135}  (September 2014).

\bibitem{eichler14}
M.~{Eichler}, A.~{Arcones}, A.~{Kelic}, O.~{Korobkin}, K.~{Langanke},
  G.~{Martinez-Pinedo}, I.~V. {Panov}, T.~{Rauscher}, S.~{Rosswog},
  C.~{Winteler}, N.~T. {Zinner} and F.-K. {Thielemann}, {\em arXiv:1411.0974}
  (November 2014).

\bibitem{metzger10b}
B.~D. {Metzger}, G.~{Martinez-Pinedo}, S.~{Darbha}, E.~{Quataert},
  A.~{Arcones}, D.~{Kasen}, R.~{Thomas}, P.~{Nugent}, I.~V. {Panov} and N.~T.
  {Zinner}, {\em MNRAS} {\bf 406} (August 2010) 2650.

\bibitem{goriely13}
S.~{Goriely}, J.-L. {Sida}, J.-F. {Lemaitre}, S.~{Panebianco}, N.~{Dubray},
  S.~{Hilaire}, A.~{Bauswein} and H.-T. {Janka}, {\em Physical Review Letters}
  {\bf 111} (December 2013)   242502.

\bibitem{rosswog14a}
S.~{Rosswog}, O.~{Korobkin}, A.~{Arcones}, F.-K. {Thielemann} and T.~{Piran},
  {\em MNRAS} {\bf 439} (March 2014) 744.

\bibitem{caballero14}
O.~L. {Caballero}, A.~{Arcones}, I.~N. {Borzov}, K.~{Langanke} and
  G.~{Martinez-Pinedo}, {\em arXiv:1405.0210}  (May 2014).

\bibitem{moeller95}
P.~M\"oller, J.~R. Nix, W.~D. Myers and W.~J. Swiatecki, {\em At. Data Nucl.
  Data Tables} {\bf 59}  (1995)   185.

\bibitem{kurtukian14}
T.~{Kurtukian-Nieto}, J.~{Benlliure}, K.-H. {Schmidt}, L.~{Audouin},
  F.~{Becker}, B.~{Blank}, I.~N. {Borzov}, E.~{Casarejos}, F.~{Farget},
  M.~{Fernandez-Ordonez}, J.~{Giovinazzo}, D.~{Henzlova}, B.~{Jurado},
  K.~{Langanke}, G.~{Martinez-Pinedo}, J.~{Pereira} and O.~{Yordanov}, {\em
  European Physical Journal A} {\bf 50} (September 2014)   135.

\bibitem{suzuki12}
T.~{Suzuki}, T.~{Yoshida}, T.~{Kajino} and T.~{Otsuka}, {\em Phys. Rev. C} {\bf
  85} (January 2012)   015802.

\bibitem{zhi13}
Q.~{Zhi}, E.~{Caurier}, J.~J. {Cuenca-Garcia}, K.~{Langanke},
  G.~{Martinez-Pinedo} and K.~{Sieja}, {\em Phys. Rev. C} {\bf 87} (February
  2013)   025803.

\bibitem{argast04}
D.~{Argast}, M.~{Samland}, F.-K. {Thielemann} and Y.-Z. {Qian}, {\em A\&A} {\bf
  416} (March 2004) 997.

\bibitem{podsiadlowski04}
P.~{Podsiadlowski}, N.~{Langer}, A.~J.~T. {Poelarends}, S.~{Rappaport},
  A.~{Heger} and E.~{Pfahl}, {\em ApJ} {\bf 612} (September 2004) 1044.

\bibitem{piran05b}
T.~{Piran} and N.~J. {Shaviv}, {\em Physical Review Letters} {\bf 94} (February
  2005)   051102.

\bibitem{stairs06}
I.~H. {Stairs}, S.~E. {Thorsett}, R.~J. {Dewey}, M.~{Kramer} and C.~A.
  {McPhee}, {\em MNRAS} {\bf 373} (November 2006) L50.

\bibitem{kitaura06}
F.~S. {Kitaura}, H.~{Janka} and W.~{Hillebrandt}, {\em A \& A} {\bf 450} (April
  2006) 345.

\bibitem{matteucci14a}
F.~{Matteucci}, D.~{Romano}, A.~{Arcones}, O.~{Korobkin} and S.~{Rosswog}, 
{\em MNRAS} {\bf 438} (January 2014) 2177 .

\bibitem{mennekens14a}
N.~{Mennekens} and D.~{Vanbeveren}, {\em A \& A} {\bf 564} (April 2014)   A134.

\bibitem{shen14a}
S.~{Shen}, R.~{Cooke}, E.~{Ramirez-Ruiz}, P.~{Madau}, L.~{Mayer} and
  J.~{Guedes} (July 2014).

\bibitem{vandevoort14a}
F.~{van de Voort}, E.~{Quataert}, P.~F. {Hopkins}, D.~{Keres} and C.-A.
  {Faucher-Giguere}, {\em MNRAS} {\bf 447} (February 2015) 140.

\bibitem{wallner14}
A.~{Wallner}, T.~{Faestermann}, J.~{Feige}, C.~{Feldstein}, K.~{Knie},
G.~{Korschinek}, W.~{Kutschera}, A.~{Ofan}, M.~{Paul}, F.~{Quinto},
G~{Rugel} and P.~{Steier},
{\em Nature Communictions} {\bf 6} (January 2015) 5956

\bibitem{hopkins14a}
P.~F. {Hopkins}, D.~{Keres}, J.~{Onorbe}, C.-A. {Faucher-Giguere},
  E.~{Quataert}, N.~{Murray} and J.~S. {Bullock}, {\em MNRAS} {\bf 445}
  (November 2014) 581.

\bibitem{li98}
L.-X. {Li} and B.~{Paczy{\'n}ski}, {\em ApJL} {\bf 507} (November 1998) L59.

\bibitem{kulkarni05}
S.~R. {Kulkarni}, {\em arXiv:astro-ph/0510256}  (October 2005).

\bibitem{metzger10a}
B.~D. {Metzger}, A.~{Arcones}, E.~{Quataert} and G.~{Martinez-Pinedo}, {\em
  MNRAS} {\bf 402} (March 2010) 2771.

\bibitem{arnett82}
W.~D. {Arnett}, {\em ApJ} {\bf 253} (February 1982) 785.

\bibitem{pinto00}
P.~A. {Pinto} and R.~G. {Eastman}, {\em ApJ} {\bf 530} (February 2000) 757.

\bibitem{kasen13a}
D.~{Kasen}, N.~R. {Badnell} and J.~{Barnes}, {\em ApJ} {\bf 774} (September
  2013)  ~25.

\bibitem{barnes13a}
J.~{Barnes} and D.~{Kasen}, {\em ApJ} {\bf 775} (September 2013)  ~18.

\bibitem{tanaka13a}
M.~{Tanaka} and K.~{Hotokezaka}, {\em ApJ} {\bf 775} (October 2013)   113.

\bibitem{grossman14a}
D.~{Grossman}, O.~{Korobkin}, S.~{Rosswog} and T.~{Piran}, {\em MNRAS} {\bf
  439} (March 2014) 757.

\bibitem{kasen14}
D.~{Kasen}, R.~{Fernandez} and B.~{Metzger}, {\em arXiv:1411.3726}  (November
  2014).

\bibitem{piran13a}
T.~{Piran}, E.~{Nakar} and S.~{Rosswog}, {\em MNRAS} {\bf 430} (April 2013)
  2121.

\bibitem{bloom06}
J.~S. {Bloom}, J.~X. {Prochaska}, D.~{Pooley}, C.~H. {Blake}, R.~J. {Foley},
  S.~{Jha}, E.~{Ramirez-Ruiz}, J.~{Granot}, A.~V. {Filippenko},
  S.~{Sigurdsson}, A.~J. {Barth}, H.-W. {Chen}, M.~C. {Cooper}, E.~E. {Falco},
  R.~R. {Gal}, B.~F. {Gerke}, M.~D. {Gladders}, J.~E. {Greene}, J.~{Hennanwi},
  L.~C. {Ho}, K.~{Hurley}, B.~P. {Koester}, W.~{Li}, L.~{Lubin}, J.~{Newman},
  D.~A. {Perley}, G.~K. {Squires} and W.~M. {Wood-Vasey}, {\em ApJ} {\bf 638}
  (February 2006) 354.

\bibitem{berger09}
E.~{Berger}, {\em ApJ} {\bf 690} (January 2009) 231.

\bibitem{kocevski10}
D.~{Kocevski}, C.~C. {Th{\"o}ne}, E.~{Ramirez-Ruiz}, J.~S. {Bloom},
  J.~{Granot}, N.~R. {Butler}, D.~A. {Perley}, M.~{Modjaz}, W.~H. {Lee}, B.~E.
  {Cobb}, A.~J. {Levan}, N.~{Tanvir} and S.~{Covino}, {\em MNRAS} {\bf 404}
  (May 2010) 963.

\bibitem{perley09}
D.~A. {Perley}, B.~D. {Metzger}, J.~{Granot}, N.~R. {Butler}, T.~{Sakamoto},
  E.~{Ramirez-Ruiz}, A.~J. {Levan}, J.~S. {Bloom}, A.~A. {Miller}, A.~{Bunker},
  H.~{Chen} and many more, {\em ApJ} {\bf 696} (May 2009) 1871.

\bibitem{rowlinson10}
A.~{Rowlinson}, P.~T. {O Brien}, N.~R. {Tanvir} and more authors, {\em MNRAS}
  (October 2010)   1479.

\bibitem{perley12}
D.~A. {Perley}, M.~{Modjaz}, A.~N. {Morgan}, S.~B. {Cenko}, J.~S. {Bloom},
  N.~R. {Butler}, A.~V. {Filippenko} and A.~A. {Miller}, {\em ApJ} {\bf 758}
  (October 2012)   122.

\bibitem{tanvir13a}
N.~R. {Tanvir}, A.~J. {Levan}, A.~S. {Fruchter}, J.~{Hjorth}, R.~A. {Hounsell},
  K.~{Wiersema} and R.~L. {Tunnicliffe}, {\em Nature} {\bf 500} (August 2013)
  547.

\bibitem{berger13b}
E.~{Berger}, W.~{Fong} and R.~{Chornock}, {\em ApJL} {\bf 774} (September 2013)
    L23.

\bibitem{piran14a}
T.~{Piran}, O.~{Korobkin} and S.~{Rosswog}, {\em arXiv:1401.2166}  (January
  2014).

\bibitem{kisaka14}
S.~{Kisaka}, K.~{Ioka} and H.~{Takami}, {\em ArXiv e-prints, arXiv:1410.0966}
  (October 2014).

\bibitem{takami14b}
H.~{Takami}, T.~{Nozawa} and K.~{Ioka}, {\em ApJL} {\bf 789} (July 2014)  ~L6.

\bibitem{mazets81}
E.~P. {Mazets}, S.~V. {Golenetskii}, V.~N. {Ilinskii}, V.~N. {Panov}, R.~L.
  {Aptekar}, I.~A. {Gurian}, M.~P. {Proskura}, I.~A. {Sokolov}, Z.~I.
  {Sokolova} and T.~V. {Kharitonova}, {\em Astrophysics and Space Science} {\bf
  80} (November 1981) 3.

\bibitem{norris84}
J.~P. {Norris}, T.~L. {Cline}, U.~D. {Desai} and B.~J. {Teegarden}, {\em
  Nature} {\bf 308} (March 1984)   434.

\bibitem{dezalay92}
J.-P. {Dezalay}, C.~{Barat}, R.~{Talon}, R.~{Syunyaev}, O.~{Terekhov} and
  A.~{Kuznetsov}, { {Short cosmic events - A subset of classical GRBs?}}, in
  {\em American Institute of Physics Conference Series\/},  eds. W.~S.
  {Paciesas} and G.~J. {Fishman}, American Institute of Physics Conference
  Series, Vol.~265 (1992), pp. 304--309.

\bibitem{kouveliotou93}
C.~{Kouveliotou}, C.~A. {Meegan}, G.~J. {Fishman}, N.~P. {Bhat}, M.~S.
  {Briggs}, T.~M. {Koshut}, W.~S. {Paciesas} and G.~N. {Pendleton}, {\em ApJL}
  {\bf 413} (August 1993) L101.

\bibitem{zhang07}
B.~{Zhang}, B.-B. {Zhang}, E.-W. {Liang}, N.~{Gehrels}, D.~N. {Burrows} and
  P.~{M{\'e}sz{\'a}ros}, {\em ApJL} {\bf 655} (January 2007) L25.

\bibitem{zhang09}
B.~{Zhang}, B.-B. {Zhang}, F.~J. {Virgili}, E.-W. {Liang}, D.~A. {Kann}, X.-F.
  {Wu}, D.~{Proga}, H.-J. {Lv}, K.~{Toma}, P.~{M{\'e}sz{\'a}ros}, D.~N.
  {Burrows}, P.~W.~A. {Roming} and N.~{Gehrels}, {\em ApJ} {\bf 703} (October
  2009) 1696.

\bibitem{paciesas03}
W.~S. {Paciesas}, M.~S. {Briggs}, R.~D. {Preece} and R.~S. {Mallozzi}, {
  {Spectral Properties of Short Gamma-Ray Bursts}}, in {\em Gamma-Ray Burst and
  Afterglow Astronomy 2001: A Workshop Celebrating the First Year of the HETE
  Mission\/},  eds. G.~R. {Ricker} and R.~K. {Vanderspek}, American Institute
  of Physics Conference Series, Vol.~662 (April 2003), pp. 248--251.

\bibitem{ghirlanda09}
G.~{Ghirlanda}, L.~{Nava}, G.~{Ghisellini}, A.~{Celotti} and C.~{Firmani}, {\em
  A \& A} {\bf 496} (March 2009) 585.

\bibitem{ghirlanda11}
G.~{Ghirlanda}, G.~{Ghisellini} and L.~{Nava}, {\em MNRAS} {\bf 418} (November
  2011) L109.

\bibitem{blinnikov84}
S.~I. {Blinnikov}, I.~D. {Novikov}, T.~V. {Perevodchikova} and A.~G.
  {Polnarev}, {\em Soviet Astronomy Letters} {\bf 10} (April 1984) 177.

\bibitem{paczynski86}
B.~{Paczynski}, {\em ApJL} {\bf 308} (September 1986) L43.

\bibitem{goodman86}
J.~{Goodman}, {\em ApJL} {\bf 308} (September 1986) L47.

\bibitem{goodman87}
J.~{Goodman}, A.~{Dar} and S.~{Nussinov}, {\em ApJL} {\bf 314} (March 1987) L7.

\bibitem{narayan92}
R.~{Narayan}, B.~{Paczynski} and T.~{Piran}, {\em ApJ} {\bf 395} (August 1992)
  L83.

\bibitem{gehrels05}
N.~{Gehrels}, C.~L. {Sarazin}, P.~T. {O'Brien}, B.~{Zhang}, L.~{Barbier}, S.~D.
  {Barthelmy}, A.~{Blustin}, D.~N. {Burrows}, J.~{Cannizzo}, J.~R. {Cummings},
  M.~{Goad}, S.~T. {Holland}, C.~P. {Hurkett}, J.~A. {Kennea}, A.~{Levan},
  C.~B. {Markwardt}, K.~O. {Mason}, P.~{Meszaros}, M.~{Page}, D.~M. {Palmer},
  E.~{Rol}, T.~{Sakamoto}, R.~{Willingale}, L.~{Angelini}, A.~{Beardmore},
  P.~T. {Boyd}, A.~{Breeveld}, S.~{Campana}, M.~M. {Chester}, G.~{Chincarini},
  L.~R. {Cominsky}, G.~{Cusumano}, M.~{de Pasquale}, E.~E. {Fenimore},
  P.~{Giommi}, C.~{Gronwall}, D.~{Grupe}, J.~E. {Hill}, D.~{Hinshaw},
  J.~{Hjorth}, D.~{Hullinger}, K.~C. {Hurley}, S.~{Klose}, S.~{Kobayashi},
  C.~{Kouveliotou}, H.~A. {Krimm}, V.~{Mangano}, F.~E. {Marshall},
  K.~{McGowan}, A.~{Moretti}, R.~F. {Mushotzky}, K.~{Nakazawa}, J.~P. {Norris},
  J.~A. {Nousek}, J.~P. {Osborne}, K.~{Page}, A.~M. {Parsons}, S.~{Patel},
  M.~{Perri}, T.~{Poole}, P.~{Romano}, P.~W.~A. {Roming}, S.~{Rosen},
  G.~{Sato}, P.~{Schady}, A.~P. {Smale}, J.~{Sollerman}, R.~{Starling},
  M.~{Still}, M.~{Suzuki}, G.~{Tagliaferri}, T.~{Takahashi}, M.~{Tashiro},
  J.~{Tueller}, A.~A. {Wells}, N.~E. {White} and R.~A.~M.~J. {Wijers}, {\em
  Nature} {\bf 437} (October 2005) 851.

\bibitem{hjorth05}
J.~{Hjorth}, D.~{Watson}, J.~P.~U. {Fynbo}, P.~A. {Price}, B.~L. {Jensen},
  U.~G. {J{\o}rgensen}, D.~{Kubas}, J.~{Gorosabel}, P.~{Jakobsson},
  J.~{Sollerman}, K.~{Pedersen} and C.~{Kouveliotou}, {\em Nature} {\bf 437}
  (October 2005) 859.

\bibitem{prochaska06}
J.~X. {Prochaska}, J.~S. {Bloom}, H.-W. {Chen}, R.~J. {Foley}, D.~A. {Perley},
  E.~{Ramirez-Ruiz}, J.~{Granot}, W.~H. {Lee}, D.~{Pooley}, K.~{Alatalo},
  K.~{Hurley}, M.~C. {Cooper}, A.~K. {Dupree}, B.~F. {Gerke}, B.~M.~S.
  {Hansen}, J.~S. {Kalirai}, J.~A. {Newman}, R.~M. {Rich}, H.~{Richer}, S.~A.
  {Stanford}, D.~{Stern} and W.~J.~M. {van Breugel}, {\em ApJ} {\bf 642} (May
  2006) 989.

\bibitem{fong13}
W.~{Fong}, W., E. {Berger}, R. {Chornock}, R. {Margutti}, A.J. 
{Levan}, N. {Tanvir}, R. {Tunnicliffe}, I. {Czekala}, D. {Fox}, D. {Perley}, 
S.B. {Cenko}, B.A. {Zauderer}, T. {Laskar}, T. S.E. {Persson}, A.J.
{Monson}, D.D. {Kelson}, C. {Birk}, D. {Murphy}, M. {Servillat}, 
G. {Anglada}, {\em ApJ} {\bf 769} (May 2013) 56.

\bibitem{berger14a}
E.~{Berger}, {\em Annual Rev. A \& A} {\bf 52} (August 2014) 43.

\bibitem{zhang04}
B.~{Zhang} and P.~{M{\'e}sz{\'a}ros}, {\em International Journal of Modern
  Physics A} {\bf 19}  (2004) 2385.

\bibitem{piran05a}
T.~Piran, {\em Reviews of Modern Physics} {\bf 76}  (2005)   1143.

\bibitem{meszaros06}
P.~{Meszaros}, {\em Reports of Progress in Physics} {\bf 69}  (2006) 2259.

\bibitem{lyutikov06}
M.~{Lyutikov}, {\em New Journal of Physics} {\bf 8} (July 2006)   119.

\bibitem{lee07}
W.~H. {Lee} and E.~{Ramirez-Ruiz}, {\em New Journal of Physics} {\bf 9}
  (January 2007)  ~17.

\bibitem{nakar07}
E.~{Nakar}, {\em Phys. Rep.} {\bf 442} (April 2007) 166.

\bibitem{gehrels09}
N.~{Gehrels}, E.~{Ramirez-Ruiz} and D.~B. {Fox}, {\em Annual Review of
Astronomy and Astrophysics} {\bf 47} (September 2009) 567.

\bibitem{kumar14}
P.~{Kumar} and B.~{Zhang}, {\em arXiv:1410.0679}  (October 2014).

\bibitem{berger07}
E.~{Berger}, {\em ApJ} {\bf 670} (December 2007) 1254.

\bibitem{nysewander09}
M.~{Nysewander}, A.~S. {Fruchter} and A.~{Pe'er}, {\em ApJ} {\bf 701} (August
  2009) 824.

\bibitem{berger11}
E.~{Berger}, {\em New Astronomy Reviews} {\bf 55} (January 2011) 1.

\bibitem{fong14}
W.~{Fong}, E.~{Berger}, B.~D. {Metzger}, R.~{Margutti}, R.~{Chornock},
  G.~{Migliori}, R.~J. {Foley}, B.~A. {Zauderer}, R.~{Lunnan}, T.~{Laskar},
  S.~J. {Desch}, K.~J. {Meech}, S.~{Sonnett}, C.~{Dickey}, A.~{Hedlund} and
  P.~{Harding}, {\em ApJ} {\bf 780} (January 2014)   118.

\bibitem{fruchter06}
A.~S. {Fruchter}, A.~J. {Levan}, L.~{Strolger}, P.~M. {Vreeswijk}, S.~E.
  {Thorsett}, D.~{Bersier}, I.~{Burud}, J.~M. {Castro Cer{\'o}n}, A.~J.
  {Castro-Tirado}, C.~{Conselice}, T.~{Dahlen}, H.~C. {Ferguson}, J.~P.~U.
  {Fynbo}, P.~M. {Garnavich}, R.~A. {Gibbons}, J.~{Gorosabel}, T.~R. {Gull},
  J.~{Hjorth}, S.~T. {Holland}, C.~{Kouveliotou}, Z.~{Levay}, M.~{Livio}, M.~R.
  {Metzger}, P.~E. {Nugent}, L.~{Petro}, E.~{Pian}, J.~E. {Rhoads}, A.~G.
  {Riess}, K.~C. {Sahu}, A.~{Smette}, N.~R. {Tanvir}, R.~A.~M.~J. {Wijers} and
  S.~E. {Woosley}, {\em Nature} {\bf 441} (May 2006) 463.

\bibitem{svensson10}
K.~M. {Svensson}, A.~J. {Levan}, N.~R. {Tanvir}, A.~S. {Fruchter} and L.-G.
  {Strolger}, {\em MNRAS} {\bf 405} (June 2010) 57.

\bibitem{bloom99}
J.~S. {Bloom}, S.~{Sigurdsson} and O.~R. {Pols}, {\em MNRAS} {\bf 305} (May
  1999) 763.

\bibitem{belczynski06}
K.~{Belczynski}, R.~{Perna}, T.~{Bulik}, V.~{Kalogera}, N.~{Ivanova} and D.~Q.
  {Lamb}, {\em ApJ} {\bf 648} (September 2006) 1110.

\bibitem{fong10}
W.~{Fong}, E.~{Berger} and D.~B. {Fox}, {\em ApJ} {\bf 708} (January 2010) 9.

\bibitem{church11}
R.~P. {Church}, A.~J. {Levan}, M.~B. {Davies} and N.~{Tanvir}, {\em MNRAS} {\bf
  413} (May 2011) 2004.

\bibitem{fryer97}
C.~Fryer and V.~Kalogera, {\em ApJ} {\bf 489}  (1997)   244.

\bibitem{fryer98}
C.~Fryer, A.~Burrows and W.~Benz, {\em ApJ} {\bf 496}  (1998)   333.

\bibitem{wang06}
C.~{Wang}, D.~{Lai} and J.~L. {Han}, {\em ApJ} {\bf 639} (March 2006) 1007.

\bibitem{wong10}
T.-W. {Wong}, B.~{Willems} and V.~{Kalogera}, {\em ApJ} {\bf 721} (October
  2010) 1689.

\bibitem{sari99}
R.~{Sari}, T.~{Piran} and J.~P. {Halpern}, {\em ApJL} {\bf 519} (July 1999)
  L17.

\bibitem{rhoads99}
J.~E. {Rhoads}, {\em ApJ} {\bf 525} (November 1999) 737.

\bibitem{panaitescu05}
A.~{Panaitescu}, {\em MNRAS} {\bf 363} (November 2005) 1409.

\bibitem{meszaros99}
P.~{Meszaros}, M.~J. {Rees} and R.~A.~M.~J. {Wijers}, {\em New Astronomy} {\bf
  4} (July 1999) 303.

\bibitem{panaitescu99}
A.~{Panaitescu} and P.~{M{\'e}sz{\'a}ros}, {\em ApJ} {\bf 526} (December 1999)
  707.

\bibitem{kumar03}
P.~{Kumar} and J.~{Granot}, {\em ApJ} {\bf 591} (July 2003) 1075.

\bibitem{granot12}
J.~{Granot} and T.~{Piran}, {\em MNRAS} {\bf 421} (March 2012) 570.

\bibitem{rybicki79}
G.~B. {Rybicki} and A.~P. {Lightman}, {\em {Radiative processes in
  astrophysics}} 1979.

\bibitem{aloy05}
M.~A. {Aloy}, H.-T. {Janka} and E.~{M{\"u}ller}, {\em A\&A} {\bf 436} (June
  2005) 273.

\bibitem{nagakura14}
H.~{Nagakura}, K.~{Hotokezaka}, Y.~{Sekiguchi}, M.~{Shibata} and K.~{Ioka},
  {\em ApJL} {\bf 784} (April 2014)   L28.

\bibitem{ruderman75}
M.~{Ruderman}, { {Theories of gamma-ray bursts}}, in {\em Seventh Texas
  Symposium on Relativistic Astrophysics\/},  eds. P.~G. {Bergman}, E.~J.
  {Fenyves} and L.~{Motz}, Annals of the New York Academy of Sciences, Vol.~262
  (October 1975), pp. 164--180.

\bibitem{schmidt78}
W.~K.~H. {Schmidt}, {\em Nature} {\bf 271} (February 1978) 525.

\bibitem{frail97}
D.~A. {Frail}, S.~R. {Kulkarni}, L.~{Nicastro}, M.~{Feroci} and G.~B. {Taylor},
  {\em Nature} {\bf 389} (September 1997) 261.

\bibitem{lithwick01}
Y.~{Lithwick} and R.~{Sari}, {\em ApJ} {\bf 555} (July 2001) 540.

\bibitem{blandford76}
R.~D. {Blandford} and C.~F. {McKee}, {\em Physics of Fluids} {\bf 19} (August
  1976) 1130.

\bibitem{norris10}
J.~P. {Norris}, N.~{Gehrels} and J.~D. {Scargle}, {\em ApJ} {\bf 717} (July
  2010) 411.

\bibitem{lazzati01}
D.~{Lazzati}, E.~{Ramirez-Ruiz} and G.~{Ghisellini}, {\em A \& A} {\bf 379}
  (December 2001) L39.

\bibitem{connaughton02}
V.~{Connaughton}, {\em ApJ} {\bf 567} (March 2002) 1028.

\bibitem{frederiks04}
D.~D. {Frederiks}, R.~L. {Aptekar}, S.~V. {Golenetskii}, V.~N. {Il'Inskii},
  E.~P. {Mazets}, V.~D. {Palshin} and T.~L. {Cline}, { {Early Hard X-ray
  Afterglows of Short GRBs with Konus Experiments}}, in {\em Gamma-Ray Bursts
  in the Afterglow Era\/},  eds. M.~{Feroci}, F.~{Frontera}, N.~{Masetti} and
  L.~{Piro}, Astronomical Society of the Pacific Conference Series, Vol.~312
  (June 2004), p. 197.

\bibitem{villasenor05}
{Villasenor}, J.~S. and {Lamb}, D.~Q. and {Ricker}, G.~R. and 
{Atteia}, J.-L. and {Kawai}, N. and {Butler}, N. and {Nakagawa}, Y. and 
{Jernigan}, J.~G. and {Boer}, M. and {Crew}, G.~B. and {Donaghy}, T.~Q. and 
{Doty}, J. and {Fenimore}, E.~E. and {Galassi}, M. and {Graziani}, C. and 
{Hurley}, K. and {Levine}, A. and {Martel}, F. and {Matsuoka}, M. and 
{Olive}, J.-F. and {Prigozhin}, G. and {Sakamoto}, T. and {Shirasaki}, Y. and 
{Suzuki}, M. and {Tamagawa}, T. and {Vanderspek}, R. and {Woosley}, S.~E. and 
{Yoshida}, A. and {Braga}, J. and {Manchanda}, R. and {Pizzichini}, G. and 
{Takagishi}, K. and {Yamauchi}, M., {\em Nature} {\bf 437} (2005) 855.

\bibitem{barthelmy05}
S.~D. {Barthelmy}, G.~{Chincarini}, D.~N. {Burrows}, N.~{Gehrels}, S.~{Covino},
  A.~{Moretti}, P.~{Romano}, P.~T. {O'Brien}, C.~L. {Sarazin},
  C.~{Kouveliotou}, M.~{Goad}, S.~{Vaughan}, G.~{Tagliaferri}, B.~{Zhang},
  L.~A. {Antonelli}, S.~{Campana}, J.~R. {Cummings} and more, {\em Nature} {\bf
  438} (December 2005) 994.

\bibitem{burrows05}
D.~N. {Burrows} and {et al.}, {\em Science} {\bf 309} (September 2005) 1833.

\bibitem{nousek06}
J.~A. {Nousek} and {et al.}, {\em ApJ} {\bf 642} (May 2006) 389.

\bibitem{chincarini07}
G.~{Chincarini}, A.~{Moretti}, P.~{Romano}, A.~D. {Falcone}, D.~{Morris},
  J.~{Racusin}, S.~{Campana}, S.~{Covino}, C.~{Guidorzi}, G.~{Tagliaferri},
  D.~N. {Burrows}, C.~{Pagani}, M.~{Stroh}, D.~{Grupe}, M.~{Capalbi},
  G.~{Cusumano}, N.~{Gehrels}, P.~{Giommi}, V.~{La Parola}, V.~{Mangano},
  T.~{Mineo}, J.~A. {Nousek}, P.~T. {O'Brien}, K.~L. {Page}, M.~{Perri},
  E.~{Troja}, R.~{Willingale} and B.~{Zhang}, {\em ApJ} {\bf 671} (December
  2007) 1903.

\bibitem{falcone07}
A.~D. {Falcone}, D.~{Morris}, J.~{Racusin}, G.~{Chincarini}, A.~{Moretti},
  P.~{Romano}, D.~N. {Burrows}, C.~{Pagani}, M.~{Stroh}, D.~{Grupe},
  S.~{Campana}, S.~{Covino}, G.~{Tagliaferri}, R.~{Willingale} and
  N.~{Gehrels}, {\em ApJ} {\bf 671} (December 2007) 1921.

\bibitem{margutti10}
R.~{Margutti}, C.~{Guidorzi}, G.~{Chincarini}, M.~G. {Bernardini}, F.~{Genet},
  J.~{Mao} and F.~{Pasotti}, {\em MNRAS} {\bf 406} (August 2010) 2149.

\bibitem{chincarini11}
G.~{Chincarini} and R.~{Margutti}, {\em International Journal of Modern Physics
  D} {\bf 20}  (2011) 1733.

\bibitem{margutti11}
R.~{Margutti}, G.~{Chincarini}, J.~{Granot}, C.~{Guidorzi}, E.~{Berger}, M.~G.
  {Bernardini}, N.~{Gehrels}, A.~M. {Soderberg}, M.~{Stamatikos} and
  E.~{Zaninoni}, {\em MNRAS} {\bf 417} (November 2011) 2144.

\bibitem{rosswog07a}
S.~Rosswog, {\em MNRAS} {\bf 376}  (2007)   L48.

\bibitem{faber06b}
J.~A. {Faber}, T.~W. {Baumgarte}, S.~L. {Shapiro} and K.~{Taniguchi}, {\em
  ApJL} {\bf 641} (April 2006) L93.

\bibitem{east12a}
W.~E. {East} and F.~{Pretorius}, {\em ApJL} {\bf 760} (November 2012)  ~L4.

\bibitem{east12b}
W.~E. {East}, F.~{Pretorius} and B.~C. {Stephens}, {\em Phys. Rev. D} {\bf 85}
  (June 2012)   124009.

\bibitem{narayan91}
R.~{Narayan}, T.~{Piran} and A.~{Shemi}, {\em ApJL} {\bf 379} (September 1991)
  L17.

\bibitem{mochkovitch93}
R.~{Mochkovitch}, M.~{Hernanz}, J.~{Isern} and X.~{Martin}, {\em Nature} {\bf
  361} (January 1993) 236.

\bibitem{popham99}
S.~W. R.~Popham and C.~Fryer, {\em ApJ} {\bf 518}  (199)   356.

\bibitem{fryer03}
C.~L. {Fryer} and P.~{M{\'e}sz{\'a}ros}, {\em ApJL} {\bf 588} (May 2003) L25.

\bibitem{birkl07}
R.~{Birkl}, M.~A. {Aloy}, H.-T. {Janka} and E.~{M{\"u}ller}, {\em A \& A} {\bf
  463} (February 2007) 51.

\bibitem{zalamea11}
I.~{Zalamea} and A.~M. {Beloborodov}, {\em MNRAS} {\bf 410} (February 2011)
  2302.

\bibitem{cooperstein87a}
J.~{Cooperstein}, L.~J. {van den Horn} and E.~{Baron}, {\em ApJL} {\bf 321}
  (October 1987) L129.

\bibitem{murguia14}
A.~{Murguia-Berthier}, G.~{Montes}, E.~{Ramirez-Ruiz}, F.~{De Colle} and W.~H.
  {Lee}, {\em ApJL} {\bf 788} (June 2014)  ~L8.

\bibitem{blandford77}
R.~D. {Blandford} and R.~L. {Znajek}, {\em MNRAS} {\bf 179} (May 1977) 433.

\bibitem{kluzniak98a}
W.~{Kluzniak} and M.~{Ruderman}, {\em ApJl} {\bf 505} (October 1998) L113.

\bibitem{usov92}
V.~V. {Usov}, {\em Nature} {\bf 357} (June 1992) 472.

\bibitem{duncan92}
R.~C. Duncan and C.~Thompson, {\em ApJL} {\bf 392} (June 1992) L9.

\bibitem{dai06}
Z.~G. {Dai}, X.~Y. {Wang}, X.~F. {Wu} and B.~{Zhang}, {\em Science} {\bf 311}
  (February 2006) 1127.

\bibitem{anderson08b}
M.~{Anderson}, E.~W. {Hirschmann}, L.~{Lehner}, S.~L. {Liebling}, P.~M. {Motl},
  D.~{Neilsen}, C.~{Palenzuela} and J.~E. {Tohline}, {\em Physical Review
  Letters} {\bf 100} (May 2008)   191101.

\bibitem{rezzolla11}
L.~{Rezzolla}, B.~{Giacomazzo}, L.~{Baiotti}, J.~{Granot}, C.~{Kouveliotou} and
  M.~A. {Aloy}, {\em ApJL} {\bf 732} (May 2011)  ~L6.

\bibitem{zrake13}
J.~{Zrake} and A.~I. {MacFadyen}, {\em \apjl} {\bf 769} (June 2013)   L29.

\bibitem{giacomazzo14}
B.~{Giacomazzo}, J.~{Zrake}, P.~{Duffell}, A.~I. {MacFadyen} and R.~{Perna},
  {\em arXiv:1410.0013}  (September 2014).

\bibitem{foucart12}
F.~{Foucart}, {\em Phys. Rev. D} {\bf 86} (December 2012)   124007.

\bibitem{stone13}
N.~{Stone}, A.~{Loeb} and E.~{Berger}, {\em Phys. Rev. D} {\bf 87} (April 2013)
    084053.

\bibitem{mcclintock14}
J.E.~{McClintock}, R.~{Narayan} and F.~F.~{Steiner}, {\em Space Science Reviews}
{\bf 183} (September 2014) 295

\bibitem{paczynski71}
B.~{Paczy{\'n}ski}, {\em Annual Review of Astronomy and Astrophysics} {\bf 9}
  (1971)   183.

\bibitem{taylor89}
J.~H. {Taylor} and J.~M. {Weisberg}, {\em \apj} {\bf 345} (October 1989) 434.

\bibitem{ruffert98}
M.~{Ruffert} and H.~{Janka}, {\em A \& A} {\bf 338} (October 1998) 535.

\bibitem{kocsis06a}
B.~{Kocsis}, M.~E. {G{\'a}sp{\'a}r} and S.~{M{\'a}rka}, {\em ApJ} {\bf 648}
  (September 2006) 411.

\bibitem{oleary09}
R.~M. {O'Leary}, B.~{Kocsis} and A.~{Loeb}, {\em MNRAS} {\bf 395} (June 2009)
  2127.

\bibitem{lee10a}
W.~H. {Lee}, E.~{Ramirez-Ruiz} and G.~{van de Ven}, {\em ApJ} {\bf 720}
  (September 2010) 953.

\bibitem{kocsis12}
B.~{Kocsis} and J.~{Levin}, {\em Phys. Rev. D} {\bf 85} (June 2012)   123005.

\bibitem{gold12}
R.~{Gold}, S.~{Bernuzzi}, M.~{Thierfelder}, B.~{Br{\"u}gmann} and
  F.~{Pretorius}, {\em Phys. Rev. D} {\bf 86} (December 2012)   121501.

\bibitem{gold13}
R.~{Gold} and B.~{Br{\"u}gmann}, {\em Phys. Rev. D} {\bf 88} (September 2013)
  064051.

\bibitem{moldenhauer14}
N.~{Moldenhauer}, C.~M. {Markakis}, N.~K. {Johnson-McDaniel}, W.~{Tichy} and
  B.~{Br{\"u}gmann}, {\em Phys. Rev. D} {\bf 90} (October 2014)   084043.

\bibitem{obergaulinger10}
M.~{Obergaulinger}, M.~A. {Aloy} and E.~{M{\"u}ller}, {\em A\&A} {\bf 515}
  (June 2010)   A30.

\bibitem{metzger08b}
B.~D. {Metzger}, E.~{Quataert} and T.~A. {Thompson}, {\em MNRAS} {\bf 385}
  (April 2008) 1455.

\bibitem{bucciantini12}
N.~{Bucciantini}, B.~D. {Metzger}, T.~A. {Thompson} and E.~{Quataert}, {\em
  MNRAS} {\bf 419} (January 2012) 1537.

\bibitem{rowlinson13}
A.~{Rowlinson}, P.~T. {O'Brien}, B.~D. {Metzger}, N.~R. {Tanvir} and A.~J.
  {Levan}, {\em MNRAS} {\bf 430} (April 2013) 1061.

\bibitem{abbott08}
B.~{Abbott}, R.~{Abbott}, R.~{Adhikari}, J.~{Agresti}, P.~{Ajith}, B.~{Allen},
  R.~{Amin}, S.~B. {Anderson}, W.~G. {Anderson}, M.~{Arain} and et~al., {\em
  ApJ} {\bf 681} (July 2008) 1419.

\bibitem{ofek08}
E.~O. {Ofek}, M.~{Muno}, R.~{Quimby}, S.~R. {Kulkarni}, H.~{Stiele},
  W.~{Pietsch}, E.~{Nakar}, A.~{Gal-Yam}, A.~{Rau}, P.~B. {Cameron}, S.~B.
  {Cenko}, M.~M. {Kasliwal}, D.~B. {Fox}, P.~{Chandra}, A.~K.~H. {Kong} and
  R.~{Barnard}, {\em \apj} {\bf 681} (July 2008) 1464.

\bibitem{hurley10}
K.~{Hurley}, A.~{Rowlinson}, E.~{Bellm}, D.~{Perley}, I.~G. {Mitrofanov}, D.~V.
  {Golovin}, A.~S. {Kozyrev}, M.~L. {Litvak}, A.~B. {Sanin}, W.~{Boynton},
  C.~{Fellows}, K.~{Harshmann}, M.~{Ohno}, K.~{Yamaoka}, Y.~E. {Nakagawa},
  D.~M. {Smith}, T.~{Cline}, N.~R. {Tanvir}, P.~T. {O'Brien}, K.~{Wiersema},
  E.~{Rol}, A.~{Levan}, J.~{Rhoads}, A.~{Fruchter}, D.~{Bersier}, J.~J.
  {Kavelaars}, N.~{Gehrels}, H.~{Krimm}, D.~M. {Palmer}, R.~C. {Duncan},
  C.~{Wigger}, W.~{Hajdas}, J.-L. {Atteia}, G.~{Ricker}, R.~{Vanderspek},
  A.~{Rau} and A.~{von Kienlin}, {\em MNRAS} {\bf 403} (March 2010) 342.

\bibitem{abadie12a}
J.~{Abadie}, B.~P. {Abbott}, T.~D. {Abbott}, R.~{Abbott}, M.~{Abernathy},
  C.~{Adams}, R.~{Adhikari}, C.~{Affeldt}, P.~{Ajith}, B.~{Allen} and et~al.,
  {\em ApJ} {\bf 755} (August 2012)  ~2.

\bibitem{shibata00}
M.~{Shibata} and K.~{\= o}. {Ury{\= u}}, {\em Phys. Rev. D} {\bf 61} (March
  2000)   064001.

\bibitem{macfadyen05}
A.~I. {MacFadyen}, E.~{Ramirez-Ruiz} and W.~{Zhang}, {\em astro-ph/0510192}
  (October 2005).

\end{thebibliography}
